\documentclass[twocolumn,pra,superscriptaddress]{revtex4-1} 
\usepackage{amsmath,amsfonts,amssymb}
\usepackage{braket}
\usepackage{graphicx}
\usepackage{subcaption}

\usepackage{float}
\makeatletter
\let\newfloat\newfloat@ltx
\makeatother
\usepackage{algorithm}
\usepackage{algpseudocode}

\usepackage[frozencache=true,cachedir=minted-cache]{minted} 
\usepackage{tcolorbox}
\usepackage{array}

\usepackage[normalem]{ulem}
\usepackage[hidelinks]{hyperref}

\begin{document}

\title{{Automated Quantum Algorithm Design using a Domain-Specific Language}}
\author{Amy Rouillard}
\affiliation{Department of Physics, Stellenbosch University, South Africa}
\author{Matt Lourens}
\affiliation{Department of Physics, Stellenbosch University, South Africa}
\author{Francesco Petruccione}
\affiliation{Department of Physics, Stellenbosch University, South Africa}
\affiliation{School for Data Science and Computational Thinking, Stellenbosch University, Stellenbosch, South Africa} 
\affiliation{National Institute for Theoretical and Computational Sciences (NITheCS), Stellenbosch, South Africa}

\begin{abstract}

We present a computational method to automatically design the $n$-qubit realisations of quantum algorithms. Our approach leverages a domain-specific language (DSL) that enables the construction of quantum circuits via modular building blocks, making it well-suited for evolutionary search. In this DSL quantum circuits are abstracted beyond the usual gate-sequence description and scale automatically to any problem size. This enables us to learn the algorithm structure rather than a specific unitary implementation. We demonstrate our method by automatically designing three known quantum algorithms—the Quantum Fourier Transform, the Deutsch-Jozsa algorithm, and Grover's search. Remarkably, we were able to learn the general implementation of each algorithm by considering examples of circuits containing at most $5$-qubits. Our method proves robust, as it maintains performance across increasingly large search spaces. Convergence to the relevant algorithm is achieved with high probability and with moderate computational resources.

\end{abstract}

\maketitle

\section{Introduction}

Quantum computers promise exponential speed-ups, yet there are only a handful of classes of quantum algorithms that realise this. Shor~\cite{shor2003haven} provides us with two possible explanations. Firstly, quantum computers operate so differently from classical computers that our existing techniques for algorithm design and our intuitions about computation may no longer be effective. Secondly, it is possible that only a limited number of such quantum algorithms exist and that many or all of them may have already been discovered. 

Given the steadily growing Quantum Algorithm Zoo~\footnote{https://quantumalgorithmzoo.org/} we might retain some optimism with regards to the latter point. Nevertheless, developing novel protocols—especially those that generate new classes of quantum algorithms, of which Shor's factorising algorithm~\cite{shor1999polynomial} is an example—is clearly a difficult task. Computational methods could offer a way to bridge the intuition gap and facilitate the discovery process. We propose such a method that uses a bottom-up approach, where a collection of well-constructed examples is translated into an algorithm.

Our approach follows the methodology of neural architecture search (NAS) where the task of automated design is divided into three parts, search space, search strategy and performance estimation strategy~\cite{elsken2019neural,wistuba2019survey,liu2021survey,white2023neural}. Neural architecture search strategies are widely used in machine learning to improve performance in computer vision, natural language processing and time series forecasting~\cite{liu2021survey}. More recently, NAS methods have also been applied in the context of variational quantum algorithms~\cite{zhang2022differentiable}, for a recent survey see~\cite{martyniuk2024quantum}. 

Quantum algorithms, which are associated with highly structured circuits, tend to be modular and to contain repeating patterns. To exploit these facts we use a domain-specific language (DSL)—specific syntax and rules for compactly expressing quantum circuits via modular building blocks, implemented as an open source Python package~\cite{lourens2023hierarchical, lourensHierarQcalQuantumCircuit2024, lourensHierarqcalGithub2024}. In this DSL, quantum circuits are abstracted beyond the usual gate-sequence description and scale automatically with problem size. This allows us to optimise the structure of the algorithm, as opposed to the quantum circuit at the level of individual gates. Previous work on automated quantum algorithm design~\cite{almuqbil2024discovery,lin2020quantum,wan2018learning,potovcek2018multi,bang2014strategy,bang2014genetic,hutsell2007applying,surkan2002evolutionary} was limited to gate-based circuit implementations of algorithms for small circuit sizes, usually $2$-$8$ qubits. In contrast, our learnt algorithms automatically generalise to arbitrary problem size which has not been demonstrated previously. Remarkably, this was achieved by using training examples related to just three small circuit sizes, namely $2$, $3$ and $4$ qubits or $3$, $4$ and $5$ qubits.

Our method is compatible with any gradient-free optimisation procedure, specifically, we use evolutionary search~\cite{de1989genetic,schwefel1995contemporary,rechenberg1973evolutionsstrategie}. The performance of candidate algorithms is measured by how closely their behaviour matches the examples provided in the training set. In addition, the evaluation takes into account the uniformity of the performance across all examples and the complexity of the candidate algorithm. We demonstrate our methodology by rediscovering three well-known quantum algorithms, the quantum Fourier transform, the Deutsch-Jozsa algorithm and Grover's search. This shows that our method works for oracle-based problems and variational circuits. Our approach also has the necessary functionality for designing both measurement-based algorithms and ansatzes for variational quantum circuits~\cite{lourens2023hierarchical, lourens2025generatinggeneralisedgroundstateansatzes}.

\begin{figure*}[ht]
    \centering
    \begin{subfigure}{0.3\textwidth}
    \begin{tcolorbox}[colback=white, colframe=black, boxrule=0.5mm, arc=0mm, left=2mm, right=2mm, top=2mm, bottom=2mm]
    \begin{minted}{python}
deutsch_jozsa = (
    QCycle(H) 
    + QPivot(Z, "*1")
    + Oracle
    + QPivot(Z, "*1")
    + QCycle(H)
) * r


    \end{minted}
    \end{tcolorbox}
    \end{subfigure}%
    \hfill
    \begin{subfigure}{0.6\textwidth}
        \centering
        \includegraphics[width = \linewidth]{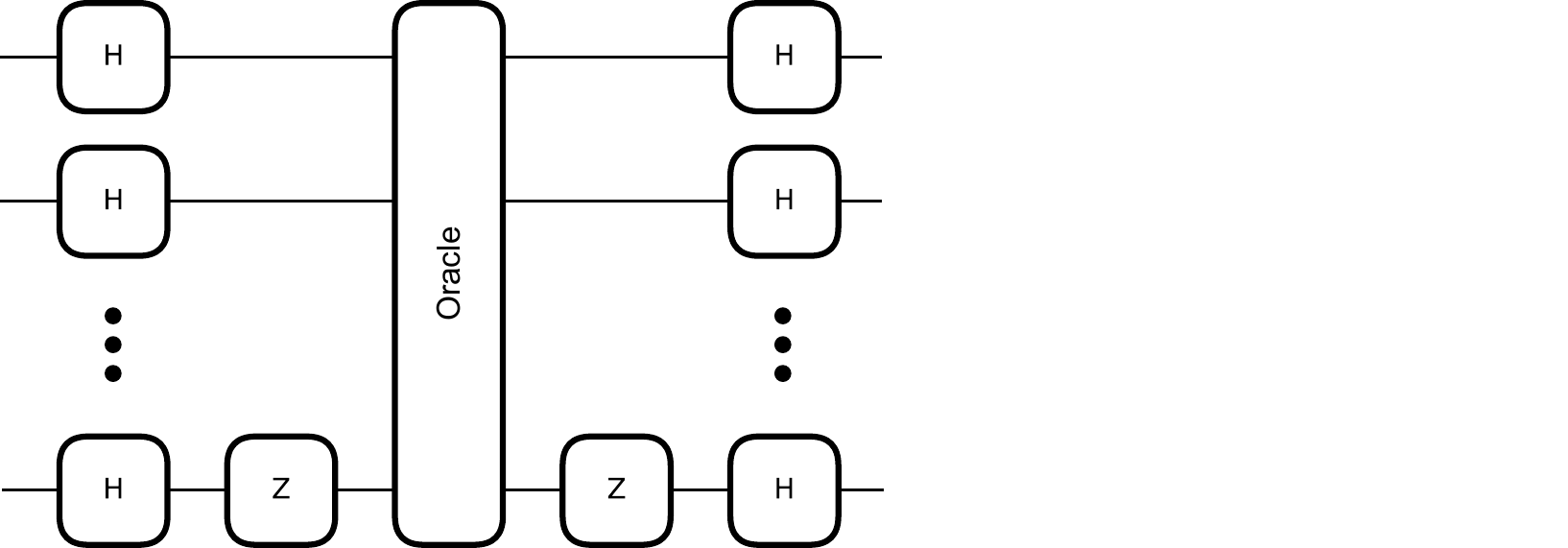}
    \end{subfigure}%
    \vspace{0.5\baselineskip}
    \begin{subfigure}{0.3\textwidth}
    \begin{tcolorbox}[colback=white, colframe=black, boxrule=0.5mm, arc=0mm, left=2mm, right=2mm, top=2mm, bottom=2mm]
    \begin{minted}{python}
grover = (
    Oracle
    + QCycle(H)
    + QCycle(X)
    + QPivot(H, "*1")
    + QPivot(MCX, "*1")
    + QPivot(H, "*1")
    + QCycle(X)
    + QCycle(H)
) * r
    \end{minted}
    \end{tcolorbox}
    \end{subfigure}%
    \hfill
    \begin{subfigure}{0.6\textwidth}
        \centering
        \includegraphics[width = \linewidth]{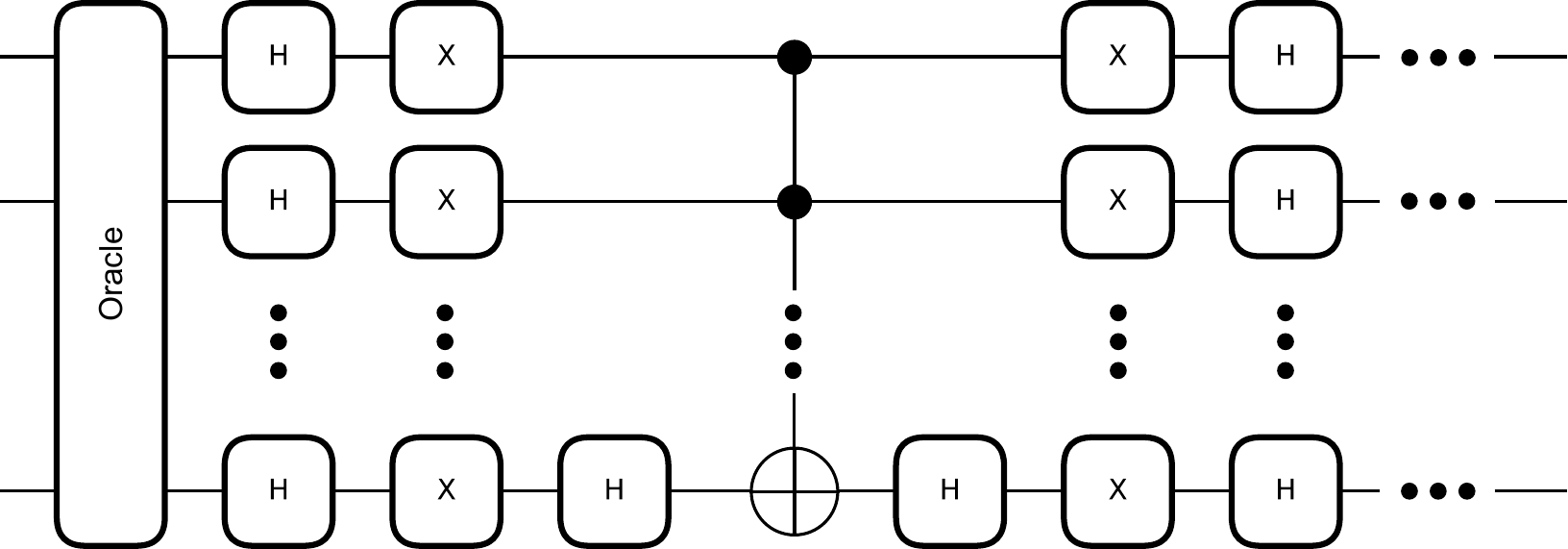}
    \end{subfigure}%
    \vspace{0.5\baselineskip}
    \begin{subfigure}{0.3\textwidth}
        \centering
        \begin{tcolorbox}[colback=white, colframe=black, boxrule=0.5mm, arc=0mm, left=2mm, right=2mm, top=2mm, bottom=2mm]
    \begin{minted}{python}
qft = (
    QPivot(H, "1*")
    + QPivot(CRphi, "1*")
    + Qmask("1*")
) * r




    \end{minted}
    \end{tcolorbox}
    \end{subfigure}%
    \hfill
    \begin{subfigure}{0.6\textwidth}
        \centering
        \includegraphics[width = \linewidth]{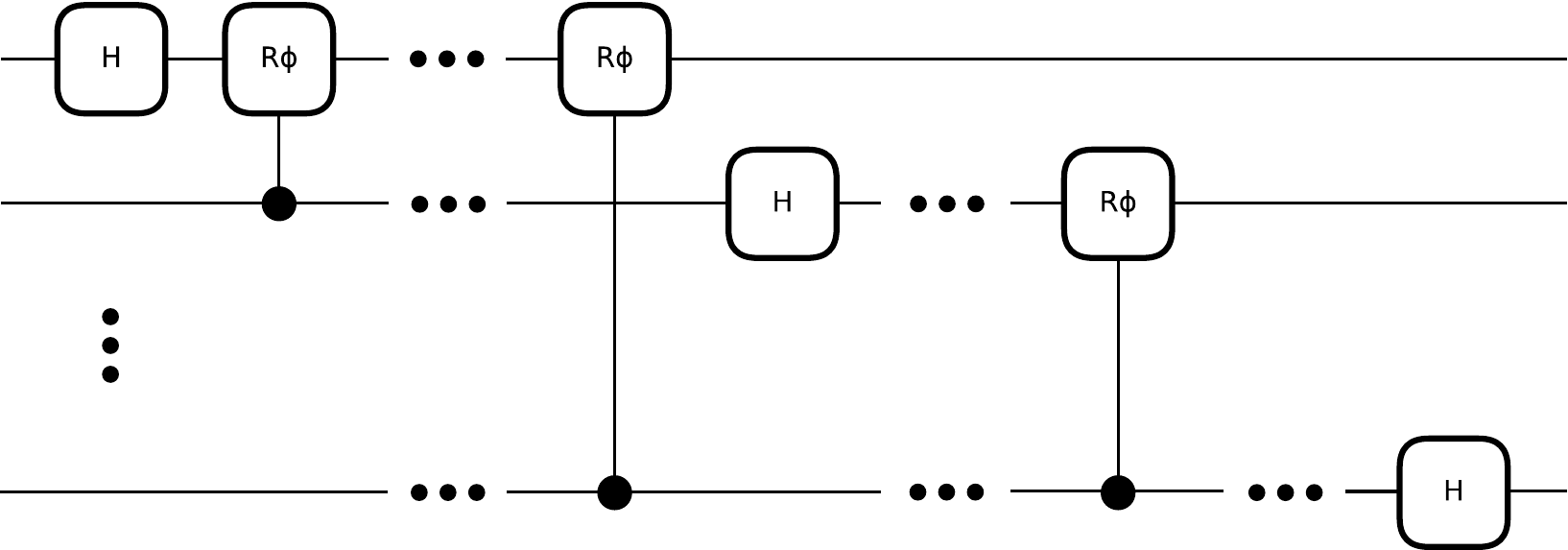}
    \end{subfigure}%
    \caption{Two equivalent $n$-qubit representations of three learnt algorithms, namely the rediscovered implementations of the Deutsch-Jozsa algorithm (top), Grover's search (middle) and the quantum Fourier transform (bottom). {On the left are DSL descriptions output by our search method, while our experiments found many variations, one example per task is shown.} {On the right} are the corresponding circuit representations that match those typically presented in the literature. The cycle (\mintinline{python}|QCycle|) and pivot (\mintinline{python}|QPivot|) motifs encode two different gate placement patterns, as can be seen by comparing the left- and right-hand sides. The mask (\mintinline{python}|QMask|) motif makes qubits unavailable.  The strings \mintinline{python}|"1*"| and \mintinline{python}|"*1"| are examples of patterns used to control the behaviour of the mask and pivot motifs, and represent the selection of the first and last qubits, respectively. The parameter \mintinline{python}|r| controls the number of times the structure is repeated, where the relationship to the circuit size can be inferred from training. 
    }\label{fig:instantiations}
\end{figure*}

\section{Results}\label{sec:results}

Using our domain-specific language combined with evolutionary search we were able to rediscover three well-known quantum algorithms, the quantum Fourier transform (QFT), the Deutsch-Jozsa algorithm and Grover's search. Figure~\ref{fig:instantiations} shows two equivalent $n$-qubit representations of three automatically designed algorithms, corresponding to each of the learning tasks. On the left-hand side of Fig.~\ref{fig:instantiations} are the leant algorithms expressed in the DSL. On the right-hand side are the corresponding quantum circuit representations, where each circuit matches the common, hardware-agnostic, implementation of each algorithm~\cite{nielsen2001quantum}. Fig.~\ref{fig:instantiations} illustrates how the DSL compactly describes the general $n$-qubit implementation of the algorithms. This allows us to immediately interpret the algorithm structure, as well as identify symmetries and patterns. For example, the symmetric structure of the diffusion term in Grover's search is immediately identifiable. 

An evolutionary algorithm is used to explore the search space where new candidate algorithms are generated by changing or combining the fittest candidates from a population of algorithms. Remarkably, we found that to converge to the general algorithm, it was sufficient to evaluate each candidate on three small circuit sizes, namely $2$, $3$ and $4$ qubits for the Deutsch-Jozsa algorithm and QFT, and $3$, $4$ and $5$ qubits for Grover's search. A candidate algorithm has a high fitness if it executes the target algorithm correctly and has low complexity, discussed further in section~\ref{sec:fitness}.  Furthermore, we did not tune the hyper-parameters of the evolutionary search. These were kept as consistent as possible across all experiments. That the same hyper-parameters were good enough for all three tasks indicates that our search strategy is stable and reliable. 

\begin{figure}[t]
    \centering
    \includegraphics[width = \columnwidth]{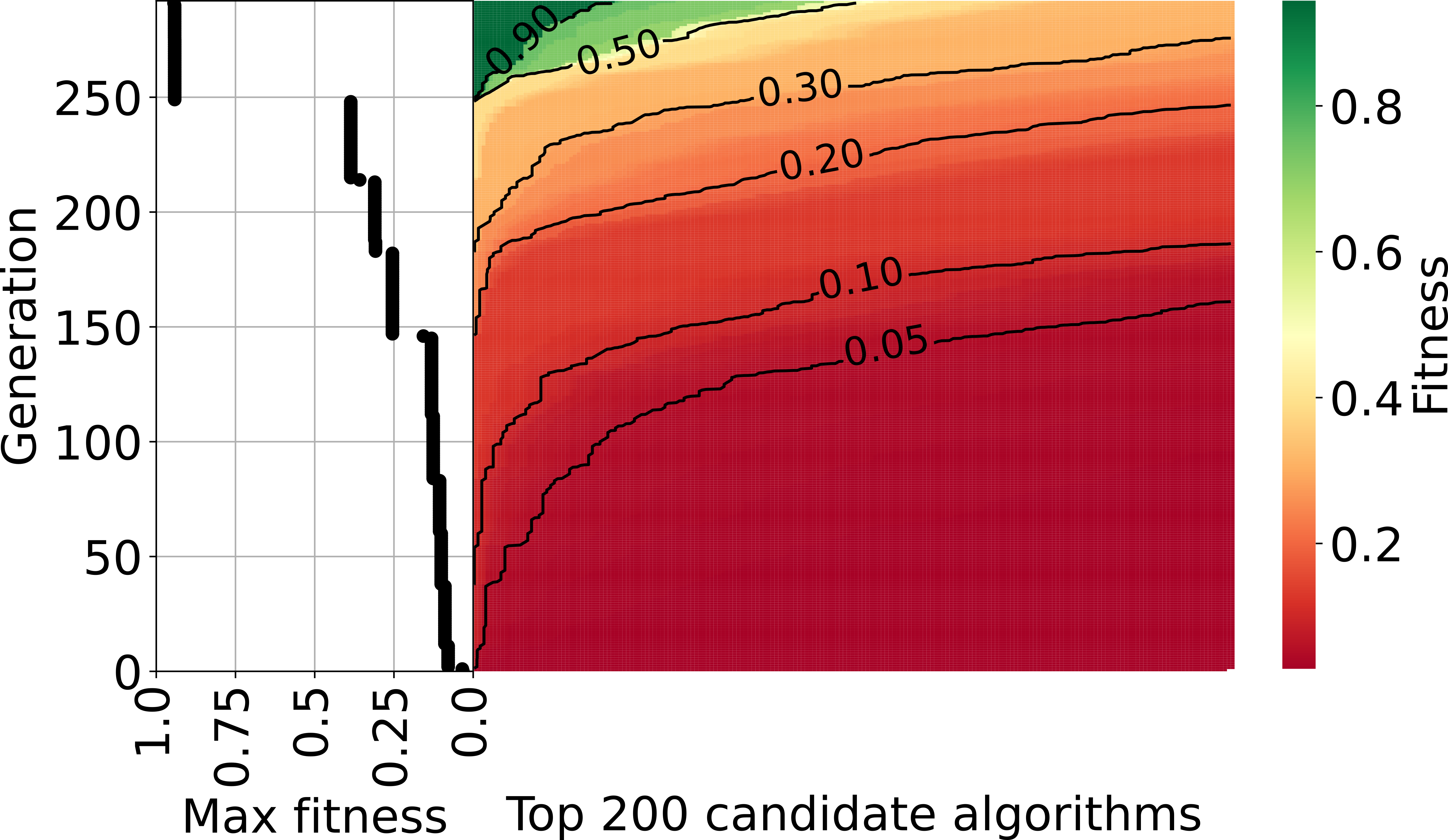}
    \caption{Evolution of the fitness scores over $292$ generations of a successful experiment in which Grover's search was rediscovered. The maximum fitness of the population at each generation of the evolutionary search (left) increases over time with periods of plateau. The heat map and contour plot (right) show the fitness of the best performing $200$ candidate algorithms sorted by fitness, decreasing from left to right. Left and right share the same $y$-axis.
    }\label{fig:eval example}
\end{figure}

\begin{figure}[t]
    \centering
    \includegraphics[width = \columnwidth]{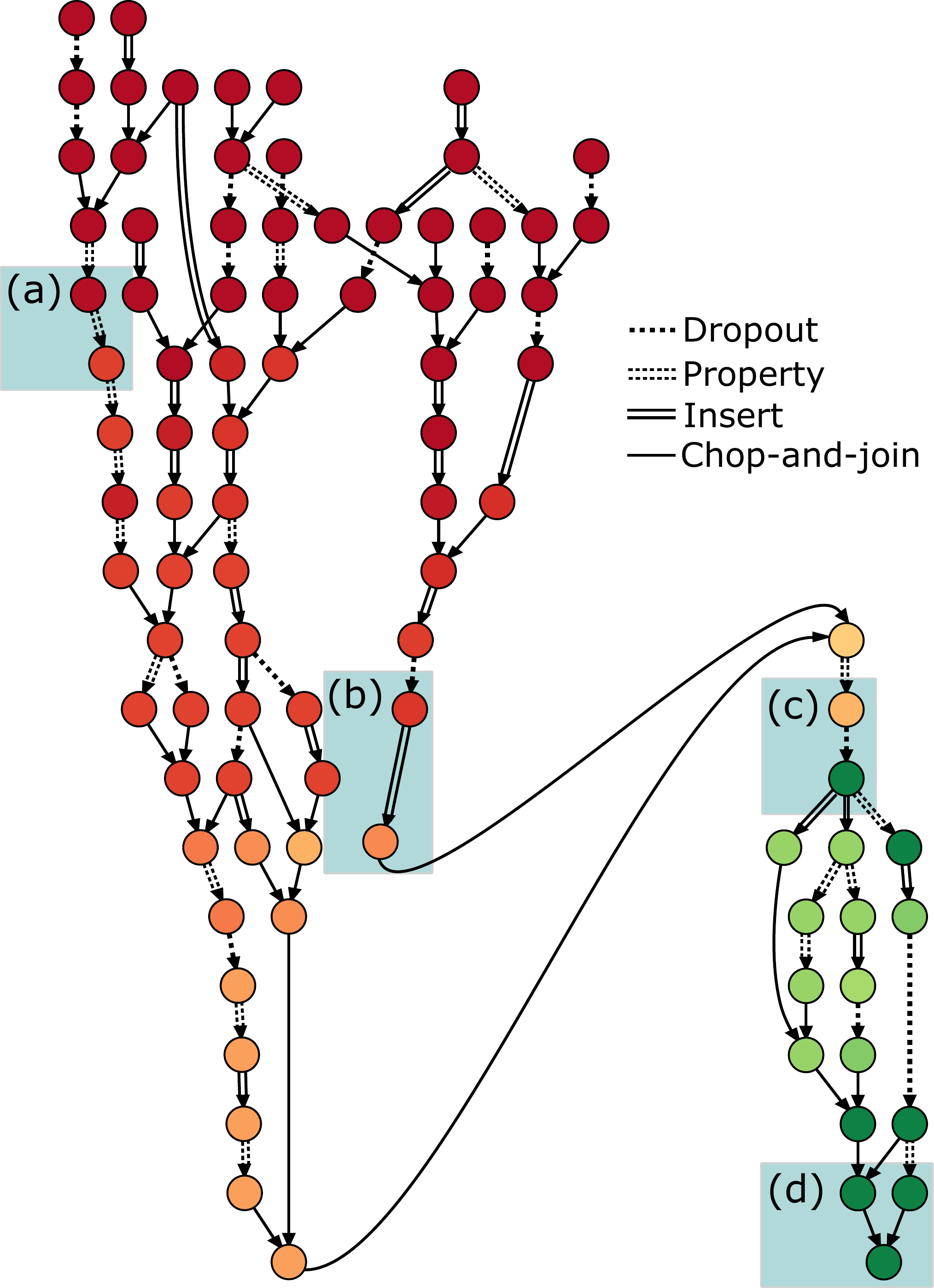}
    \caption{Directed graph showing the subset of the population that evolved into an implementation of Grover's search. Each node represents a candidate algorithm from the same population shown in Fig.~\ref{fig:eval example}. The colour of each node corresponds to the fitness, matching the scale in Fig.~\ref{fig:eval example}. Each mutation type is represented by a different edge. Four specific mutations, each of a different type, are highlighted in blue and elaborated on in Fig.~\ref{fig:mutations}.
    }\label{fig:tree}
\end{figure}

Fig.~\ref{fig:eval example} and Fig.~\ref{fig:tree} detail a successful experiment that was able to rediscover Grover's search. Fig.~\ref{fig:eval example} shows an example of the distribution of the population fitness over the course of several generations. The left panel of Fig.~\ref{fig:eval example} illustrates the typical step-wise increase of the maximum fitness over time. While fitness initially increases steadily, there is a significant jump in fitness at the $249$-th generation. This is owing to the `dropout' mutation labelled (c) in Fig.~\ref{fig:tree} and detailed in Fig.~\ref{fig:mutations}. Since candidate algorithms are defined by a collection of discrete properties, large jumps in fitness are to be expected for single discrete changes. 

The right panel of Fig.~\ref{fig:eval example} shows the changing composition of the population. The slope of the contour lines indicates the rate of displacement of the poorest performing candidates from the top $200$ algorithms. The steeper the slope of a contour line, the slower poorer-performing candidates are displaced. The general trend is that the gradient of each contour line decreases over time as the current best algorithm, its duplicates, and related variants propagate rapidly and take up an increasing proportion of the population. 

Fig.~\ref{fig:tree} shows the evolutionary path, represented as a directed graph, that led to Grover's search. The graph contains a roughly equal distribution of mutation types, represented by different edges. This indicates that all four mutation types played an important role in producing the final algorithm, which was the case for all experiments across all three tasks.  In the literature, it has been suggested that mutations that combine two candidates, in our case the `chop-and-join' mutation, do not lead to better-performing architectures~\cite{white2023neural,rattew_domain-agnostic_2020}. Our results indicate that in the context of algorithm design, this is not the case. A possible explanation is that we break each architecture into two pieces before recombining them, rather than simply appending algorithms head-to-tail. This provides an opportunity for useful sections of each algorithm to be combined. Dividing each algorithm into two randomly determined pieces before recombination also ensures that the size of candidates does not double with each application of this mutation. Furthermore, random division ensures diversity in the size of candidates. 

\begin{figure}[t]
    \centering
    \includegraphics[width = \columnwidth]{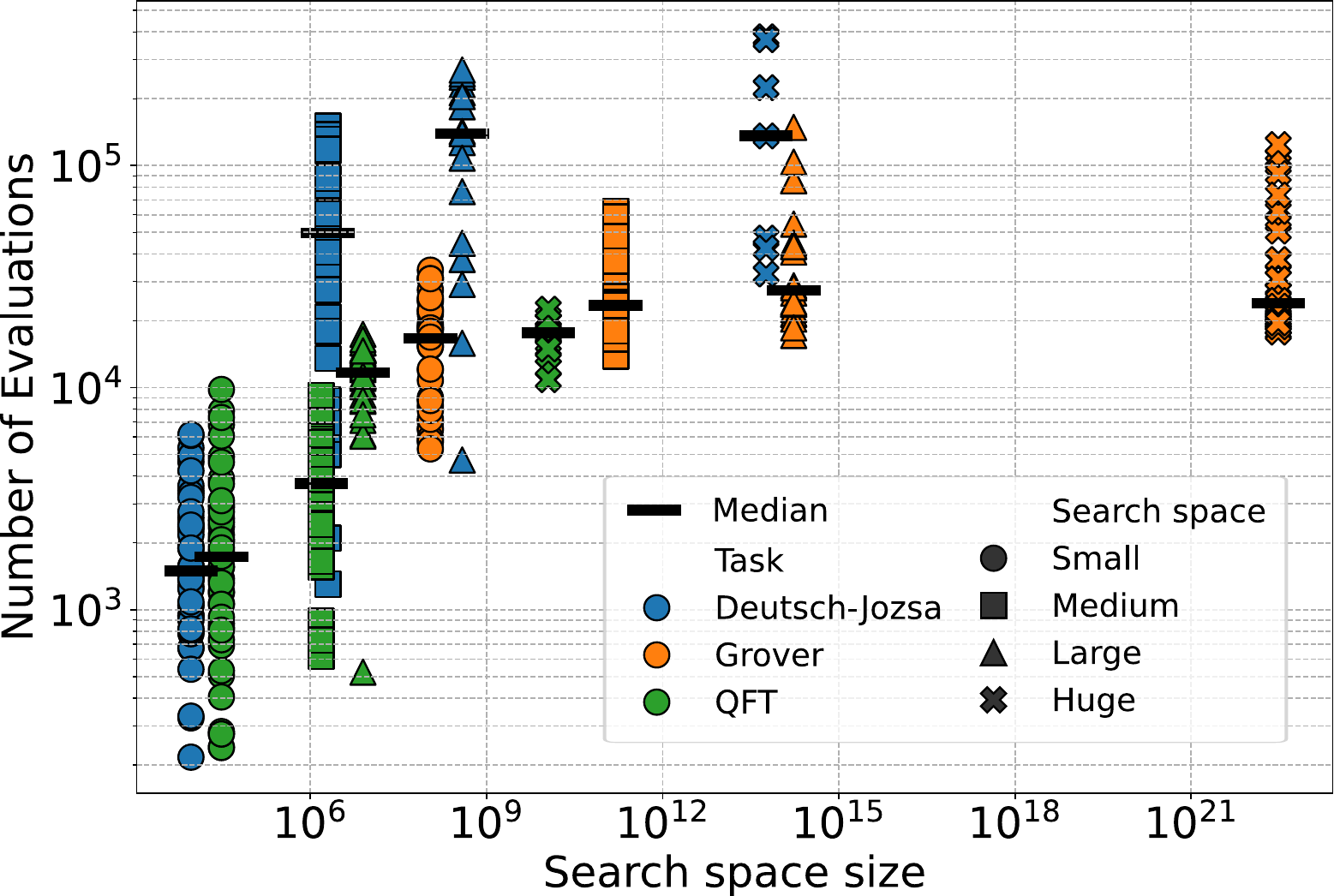}
    \caption{Number of evaluations required to successfully design three different quantum algorithms. For each task, four search spaces of increasing size were tested. The median number of evaluations is indicated with a horizontal bar.  }\label{fig:eval main}
\end{figure}

We demonstrate the robustness of our method by investigating the effect of increasing search space size.  The DSL that was used to represent our search space is implemented as an open-source Python package called Hierarqcal~\cite{lourens2023hierarchical,lourens2025generatinggeneralisedgroundstateansatzes}. See Section~\ref{sec:Hierarqcal} and Fig.~\ref{fig:instantiations} for an illustration of the basic functionality of Hierarqcal. In this framework, algorithms are built up of motifs that encode different gate placement patterns. Each motif has several properties, each of which can take on a possibly infinite number of discrete values. Motif properties include mapping, step, stride, offset, boundary condition, global and merge within patterns, weight sharing and edge order. Selecting appropriate sets of discrete values for each of these properties determines the configuration space, which in turn determines the set of possible algorithms that can be constructed, i.e. the search space.

During evolutionary search, no restrictions are placed on the number of Hierarqcal motifs used to construct an algorithm. Consequently, the size of the search space is infinite. For the purposes of analysis, we instead refer to the size of the search space as the number of algorithms that would be evaluated in a worst-case-scenario brute-force search using our DSL. A worst-case-scenario brute-force search would involve evaluating all possible algorithms composed of at most $n$ motifs, where $n$ is the minimum number of motifs known to be required to build the algorithm. This is a useful way to compare experiments and can also be used to compare performance to random search.

In Fig.~\ref{fig:eval main} we see that the number of candidates evaluated in each experiment was a small fraction of the size of the search space, indicating the efficiency of our evolutionary strategy. The number of evaluations required grows with the size of the search space, though favourably at a decreasing rate. A total of $50$ experiments were performed for each combination of task and search space size, and successful runs are shown in Fig.~\ref{fig:eval main}. An experiment is considered successful if it is completed within a certain runtime limit and if certain stopping conditions are met, which we describe in detail in Section~\ref{sec:fitness}. Table~\ref{tab:eval success rate} shows the approximate success rates for each task and notes the runtime limits in core processing unit (CPU) hours. Taking the success rates into account, our method is robust and can plausibly be run on a personal computer {-- $8$ CPUs -- with a $90\%$ chance of success within two days in most cases.} 

\begin{table}[t]
\caption{runtime limit, approximate success rate and median runtime of successful experiments for three different learning tasks and corresponding search spaces. Fifty experiments were run per configuration. }\label{tab:eval success rate}
\begin{tabular}{ |p{4em}|p{4em}|p{4.7em}<{\raggedleft\arraybackslash}|p{6em}<{\raggedleft\arraybackslash}|p{4.7em}<{\raggedleft\arraybackslash}|} 
\hline
Task & Search space  & Limit (CPU hrs) & Success rate & Median (CPU hrs) \\
\hline
Deutsch- & Small & 28 & 1.00 & 0.3   \\ 
Jozsa & Medium & 56 & 0.58 & 15.4   \\ 
 & Large & 112 & 0.34  & 42.9  \\ 
  & Huge & 112 & 0.14  & 33.7 \\ 
 \hline
Grover & small& 28 & 0.84 & 10.4 \\
 & Medium& 56 & 0.62 & 13.2  \\ 
 & Large& 112 & 0.44 & 13.7 \\ 
  & Huge & 112 & 0.60 & 10.5 \\ 
 \hline
QFT & Small& 56 & 0.94 & 8.8 \\ 
 & Medium & 112 & 0.62 & 51.8  \\ 
 & Large & 168 & 0.64 & 116.6  \\ 
  & Huge & 168 & 0.26 & 129.5  \\ 
\hline
\end{tabular}
\end{table}

When parametrised unitary gates are used, as was the case for the QFT algorithm, runtimes are longer since the parameters of these gates have to be trained during evaluation. We note that, by inspecting the learnt gate parameters it is possible to infer the general relationship between parameters, which may also depend on the circuit sizes. Similarly for all tasks, the relationship between the circuit size and number of repetitions, \mintinline{python}|r| in Fig.~\ref{fig:instantiations}, can also be inferred in this way. For example, for QFT we saw that the learnt number of repetitions was equal to the circuit size.

\section{Discussion}

{Our results demonstrate the utility of our approach as a tool for quantum algorithm design. } We successfully rediscover the quantum Fourier transform (QFT), the Deutsch-Jozsa algorithm, and Grover’s search, validating our approach {and outlining how it could be used to discover a new algorithm given a well-posed task}. {Beyond quantum algorithms, our approach is also applicable to the design of variational Ans\"atze for quantum many-body states, as demonstrated in \cite{lourens2025generatinggeneralisedgroundstateansatzes}. In \cite{lourens2025generatinggeneralisedgroundstateansatzes}, the same DSL allows for the discovery of new interpretable wavefunctions that remain analytically tractable. Applied to ground-state problems, the method autonomously rediscovers a mean-field description and extends it to incorporate correlations, producing wavefunctions that extrapolate to arbitrary system size without modification.}

Central to our method is a domain-specific language, which captures the inherent scalability of algorithms. Further, the DSL biases the algorithms in the search space to be structured, capturing modularity and repeating patterns. As a result, the learnt algorithms are immediately interpretable and analysable. Notably, convergence to the general algorithm requires evaluation on only three small circuit sizes, containing at most $5$ qubits, which significantly reduces the computational cost. 

The DSL informs the three aspects of our NAS-inspired design approach. It enables the compact representation of algorithms in the search space. It makes constructing useful mutations for evolutionary search straightforward. And, it allows us to evaluate candidate algorithms on a collection of different circuit sizes. 

Furthermore, the DSL contains a small number of building blocks that in turn depend on a limited number of properties. Therefore, we can greatly reduce the size of the search space compared to using a gate-based circuit description. Importantly, we are able to reduce the search space size without sacrificing expressivity, since the DSL captures the essential characteristics of algorithms, namely structure and scalability. 

Our search strategy is robust and efficiently explores the search space. As the search space size increases, the number of evaluations required grows at a diminishing rate, and the fraction of the space explored decreases rapidly. Good performance was achieved for all tasks, despite maintaining consistent hyper-parameters and without the use of tuning. We found that all mutation types contributed meaningfully to the search, including a mutation that combines two algorithms. 

By using a bottom-up approach, we reformulate the task of algorithm design as a task of creating a set of well-constructed training examples. Our performance estimation strategy then evaluates candidate algorithms based on their ability to generalise across these training examples. We prioritise simpler algorithms by incorporating a complexity penalty while simultaneously rewarding algorithms that perform uniformly across training examples. This, combined with the attributes mentioned above, enables us to automatically design the $n$-qubit realisation of a quantum algorithm.

\section{Methods}\label{sec:methods}

Much of our success is due to the use of a domain-specific language. It allows us to learn the general algorithm structure rather than a specific unitary implementation of the algorithm for some number of qubits. The DSL was adapted from its initial formalism detailed in \cite{lourens2023hierarchical}, where many of the core ideas were introduced. For a recent overview, see the blog post \cite{lourensHierarQcalQuantumCircuit2024}, and for the associated Python package Hierarqcal, see the GitHub repository~\cite{lourensHierarqcalGithub2024}.

The original inspiration for the DSL came from the field of neural architecture search~\cite{liu_hierarchical_2018}. In line with NAS we divide our automated algorithm design approach into three components: search space, search strategy, and performance estimation~\cite{elsken2019neural, wistuba2019survey, liu2021survey}. The DSL is used to specify the search space, see Section~\ref{sec:Hierarqcal}. Owing to the modular nature of our DSL, evolutionary search is a natural choice for search strategy, detailed in Section~\ref{sec:genetic_alg}. The performance estimation method used to evaluate candidate algorithms is described in Section~\ref{sec:fitness}. Finally, in Section~\ref{sec:tasks}, we provide an overview of each design task, with a particular focus on problem setup. 

\subsection{Search space}\label{sec:Hierarqcal}

The DSL provides a way to encode quantum circuits, in fact, it can be used to encode arbitrary compute graphs of which quantum circuits are an example. In place of representing the quantum circuit as a sequence of individual gate placements, we instead work on the abstract level of placement patterns, which we term motifs. Each motif encodes a different pattern of gate placement, see Fig.~\ref{fig:instantiations}. By design, a motif encodes the gate placements for any circuit size, where scaling happens automatically and in an intuitive way. The definitions of the motifs were heuristically motivated, and capture the regular patterns that we see throughout quantum circuits that perform structured tasks.

{The motifs are the building blocks of our DSL and are used to generate hypergraphs. We denote a hypergraph by $\mathcal{G}=(Q,E)$, where the sequences $Q$ and $E$ list qubit indices and edges, respectively. Each edge $e\in E$ encodes the connectivity of an associated unitary operation $U_{e}$. For example, a $3$-qubit unitary gate $U_{3,1,5}$ acts on qubits three, one and five, and is represented by the edge $e=(3,1,5)$ in the hypergraph.  A candidate algorithm $\mathcal{A}=(M_1,\dots, M_p)$ is a sequence of $p$ motifs. For example, the QFT, Fig.~\ref{fig:instantiations} left, is a sequence of three motifs, two consecutive pivots followed by a mask, repeated $r$ times. A candidate algorithm $\mathcal{A}$ is instantiated by specifying a list of qubit indices $Q$. Each motif $M_m$ generates the hypergraph $\mathcal{G}_m=(Q,E_m)$ with associated unitary operation $U^{(m)}$. The list of qubit indices $Q$ does not carry a subscript $m$ since the edges of all hypergraphs encode unitary operations acting on the same quantum circuit. The quantum circuit, with qubit indices $Q$, corresponding to a candidate algorithm $\mathcal{A}$ 
is given by
\begin{align}
U\left(Q,\mathcal{A}=\left(M_1,\dots, M_p\right)\right)
  &= 
      \mathop{\prod}\limits_{m\in[1,p]}^{\leftarrow}
      \ \mathop{\prod}\limits_{e\in E_m}^{\leftarrow}
      U^{(m)}_{e},
\label{eq:hypergraph_build_state}
\end{align}
where the product of unitaries is ordered with indices increasing from right to left, as indicated by the arrows. In Eq.~\eqref{eq:hypergraph_build_state}, we see that the sequence of edges $E_m$ determines the gate placement patterns within the circuit. In turn, each motif $M_m$ generates the sequence of edges $E_m$ based on a rule that captures size scaling. In this way, the algorithm $\mathcal{A}$ is built without having to specify a system size. Eq.~\eqref{eq:hypergraph_build_state} enables the analytical analysis of the resulting circuit, and allows us to verify whether the generalisation of the algorithm to any problem size is correctly captured.}

Each motif type corresponds to a different method of edge generation. {The cycle iterates through the list of nodes when creating edges, while the pivot connects one or more fixed nodes to all remaining nodes. The mask motif makes nodes unavailable to subsequent motifs, while its counterpart the unmask makes unavailable nodes available again.} The specific behaviour of each motif type is controlled by several properties that influence edge generation.  For example, the global pattern property of the pivot is used to control which nodes are fixed in the one-to-many edge generation pattern. {Appendix~\ref{app:DSL} provides further explanation of the DSL by example.}

Motifs can be composed to create a sequence of motifs, which can in turn be composed to form higher-level motifs, and so on.  When the number of nodes (qubits) is specified, the edges are generated according to the order of the motifs in the composition. For example, in the representation of the Deutsch-Jozsa algorithm in Fig.~\ref{fig:instantiations} the cycle of the Hadamard gates is placed before the pivot of the Pauli-Z gates in the circuit representation according to the order of operations in the composition of motifs shown on the left. 

For each of the tasks we tested, we imposed a fixed macro-architecture template at the highest level of the structural hierarchy. This may not be applicable in all scenarios and can easily be modified or removed. Specifically, we assume that the algorithm is made up of a composition of motifs repeated $r$ times. The number of repetitions $r$ may depend on the problem size and is a trainable parameter. We also point out that any single gate placement can be achieved using the pivot motif with an appropriate global pattern. This ensures that it is always possible to map a gate-based circuit representation into the DSL.

A quantum circuit is itself a unitary operation and can therefore be associated with a motif. This allows larger structures to be built from sub-substructures, for example building multiplication from addition. While it was not necessary to exploit this functionality in the current work, it allows structurally complex algorithms to be represented compactly. For instance, Shor's factorisation algorithm lends itself well to this. It is straightforward to allow an algorithm from the population to be associated with a motif, thereby building larger complex structures. However, we speculate that for the search to remain robust and efficient, it would be necessary to implement a more sophisticated search strategy.

A hierarchical representation of a large architecture is an effective way to reduce search complexity and leads to better search efficiency~\cite{liu_hierarchical_2018, ru_neural_2020, christoforidis_novel_2023}. By limiting the search space to a small number of motifs, that in turn depend on a limited number of properties, we can significantly reduce the size of the search space compared to a gate-sequence circuit encoding. Importantly, we are able to reduce the search space complexity without sacrificing expressivity. Algorithms are inherently scalable and our DSL captures this property. Further, our DSL biases the search space toward structured circuits which are likely to include the structured operations necessary to build algorithms. In this way, we were able to automatically design large circuit structures in reasonable times, as demonstrated in Fig.~\ref{fig:eval main} and Table~\ref{tab:eval success rate}. 

Our DSL simplifies the search strategy compared to approaches that use a gate-based circuit representation. Using the DSL, it is no longer necessary to construct a set of rules to govern gate placements within the circuit during search~\cite{rattew_domain-agnostic_2020}.  Instead, we define a configuration space, that is a set of available discrete values for each property of the motifs. This in turn determines the set of possible algorithms that can be constructed. {For example, for Grover's search algorithm and the largest configuration space, labelled \textit{Huge} in Fig.~\ref{fig:eval main}, the number of possible unique motifs is $646$, where the oracle is also counted as a motif. This results in a search space size of approximately $3\times10^{22}$, where as discussed in Sec.~\ref{sec:results} the search space size is the number of algorithms that would be evaluated in a worst-case-scenario brute-force search~\footnote{A count of all the possible candidate algorithms consisting of up to $p$ motifs is given by $646 + 646^2 +\cdots + 646^p = {646(646^p-1)}/({646-1})$. The minimum number of motifs required to represent Grover's search algorithm is eight. }. 
In contrast, for an equivalent gate-based search method, the search space size scales exponentially with circuit size. In our method, the search space size is independent of the number of qubits. } 

{In \cite{almuqbil2024discovery} the authors also simplified the search space by using a vector to encode the sequence of gates to be applied, where it was assumed that each gate in the sequence is applied to all qubits. In the case of two-qubit gates, these are applied from each qubit to the next without looping back, which is equivalent to a cycle with open boundary conditions in our DSL. Using this representation, the authors were able to rediscover Grover's search algorithm for circuits containing $3$-$8$ qubits. In their vector-based representation, the assumption that a gate acts on each qubit in the circuit was motivated by the permutation symmetry of Grover's search and therefore does not generalise to other algorithms that do not have the same permutation symmetry, including QFT and the Deutsch-Jozsa algorithm.}

For a given task, heuristically motivated constraints can be applied to the property values in the configuration space. For example, the global pattern can be restricted to values that reflect information about qubits that have a special significance. In our implementation, the oracle for the Deutsch-Jozsa algorithm performs a multi-controlled bit flip on the last qubit. This makes the global pattern that encodes a pivot on the last qubit relevant to the algorithm structure. While a pivot on a middle qubit is unlikely to be relevant since the middle qubit has no special significance. Similarly, if we interleave ancillary qubits into our circuit then we are justified in assuming that the patterns ``10" and ``01" are relevant. 

\subsection{Search strategy}\label{sec:genetic_alg}

Evolutionary search is a gradient-free optimisation technique inspired by the process of natural selection. It remains a popular optimisation method due to its flexibility, conceptual simplicity, and competitive results~\cite{white2023neural}. Beginning with a collection of randomly generated candidate algorithms, the fittest candidates are selected and used to produce the next generation, growing the population over time. New algorithms are produced by combining two algorithms or by changing a property of a single algorithm. These operations are referred to as mutations. The pseudo-code that summarises our optimisation procedure is given in Algorithm~\ref{alg:genetic}.

\begin{algorithm}
\caption{Evolutionary algorithm: An iterative optimisation process that maintains a population of candidate algorithms, refining them through selection, mutation, and evaluation. The search continues until a predefined stopping condition is met. Arrows indicate assignment and plus operations indicate appendage. }\label{alg:genetic}
\begin{algorithmic}[1]
\Procedure{EvolutionarySearch}{}
\State $population \gets InitialisePopulation()$ \label{line:init}
\State $flag \gets StoppingCondition(population)$ \label{line:stop}
\While{$\lnot flag$}
    \State $children \gets []$
    \For{$i \gets 1, n_{batch}$} \label{line:batch}
    	\State $m_1, m_2 \gets \textsc{Selection}(population)$ \label{line:selection}
	   \State $children \gets children + \textsc{Mutate}(m_1,m_2)$ \label{line:mutation}
    \EndFor
    \State $population \gets population + \textsc{Evaluate}(children)$ \label{line:eval}
    \State $flag \gets StoppingCondition(population)$
\EndWhile
\State \textbf{return} $population$
\EndProcedure
\end{algorithmic}
\end{algorithm}

We initialise our population by selecting a random sample of algorithms from the set of algorithms composed of a small number of motifs. The properties of these motifs are determined by the configuration space. In particular, we start with a sample of $200$ algorithms composed of $2$ or $3$ motifs, depending on the size of the configuration space. These algorithms are then evaluated to form the initial population of algorithms, see Alg.~\ref{alg:genetic} line~\ref{line:init}. 

In the selection step, we choose two candidate algorithms to mutate, line~\ref{line:selection} of Alg.~\ref{alg:genetic}.  Here we balance exploration and exploitation, that is trying new algorithms versus further optimising ones that already have a high fitness score compared to the rest of the population. This is done by randomly deciding between two selection methods, namely random selection or tournament selection (i.e. selection according to fitness ranking) which emphasise exploration and exploitation, respectively. Tournament selection depends on selection pressure which determines the percentage of the population considered. A higher selection pressure corresponds to greater exploitation since each candidate competes against a larger subset of the total population. The probability of performing random or tournament selection and the selection pressure are hyper-parameters of the search strategy. These can be changed over the course of an experiment. We begin in a highly exploratory phase followed by an exploitative phase, with the transition between these phases scheduled according to a sine function. By using a periodic function, we can alternate between phases of exploration and exploitation during long experiments.

\begin{figure}[t]
    \centering
    \begin{subfigure}{0.23\textwidth}
    \begin{tcolorbox}[colback=white, colframe=black, boxrule=0.5mm, arc=0mm, left=1mm, right=0mm, top=1mm, bottom=1mm, width=\textwidth, nobeforeafter]
    \begin{minted}[highlightlines={3-4}]{python}
algorithm = (
 QPivot(H, "1*") 
 + QPivot(H, "*1") # -
 + QPivot(H, "1*") # +
 + QCycle(X)
 + QPivot(MCX, "*1")
 + QPivot(MCX, "*1") 
 + QPivot(MCX, "*1")
 + Oracle
 + QCycle(H)
) * r
    \end{minted}
    \end{tcolorbox}
    \subcaption{Property mutation: A motif and a property thereof are randomly selected. Here, the global pattern of the selected pivot is changed from *1 to 1*.}
    \end{subfigure}
    \hfill
    \begin{subfigure}{0.23\textwidth}
    \begin{tcolorbox}[colback=white, colframe=black, boxrule=0.5mm, arc=0mm, left=1mm, right=0mm, top=1mm, bottom=1mm, width=\textwidth, nobeforeafter]
    \begin{minted}[highlightlines={5}]{python}
algorithm = (
 Oracle
 + QCycle(H)
 + QCycle(X)
 + QPivot(H, "*1") # +
 + QPivot(MCX, "*1")
 + QCycle(x)
) * r
    \end{minted}
    \end{tcolorbox}
    \subcaption{Insert motif: A random motif is inserted at a random position in the larger composition.}
    \end{subfigure}
    \\
    \begin{subfigure}{0.23\textwidth}
    \begin{tcolorbox}[colback=white, colframe=black, boxrule=0.4mm, arc=0mm, left=1mm, right=0mm, top=1mm, bottom=1mm, width=\textwidth, nobeforeafter]
    \begin{minted}[highlightlines={10}]{python}
algorithm = (
 Oracle
 + QCycle(H)
 + QCycle(x)
 + QPivot(H, "*1")
 + QPivot(MCX, "*1")
 + QPivot(MCX, "*1")
 + QPivot(MCX, "*1")
 + QCycle(H)
 + Oracle # -
 + Oracle
 + Oracle
 + QCycle(H)
 + QPivot(H, "*1")
 + QCycle(x)
 + QCycle(H)
) * r
    \end{minted}
    \end{tcolorbox}
    \subcaption{Dropout: A randomly selected motif is removed.}
    \end{subfigure}
    \hfill
    \begin{subfigure}{0.23\textwidth}
        \begin{tcolorbox}[colback=white, colframe=black, boxrule=0.5mm, arc=0mm, left=1mm, right=0mm, top=1mm, bottom=1mm, width=\textwidth, nobeforeafter]
    \begin{minted}[highlightlines={2-7}]{python}
algorithm = (
  # head
  Oracle            
  + QCycle(H)       
  + QCycle(x)                
  + QPivot(H, "*1")          
  + QPivot(MCX, "*1")
  # tail
  + QPivot(H, "*1")
  + QCycle(x)
  + QCycle(H)
) * r
    \end{minted}
    \end{tcolorbox}
    \subcaption{Chop and join: Two pieces, each from a different algorithm, are combined to produce a new algorithm. }
    \end{subfigure}%
    \caption{Examples of mutations, written in the DSL. Mutations (a), (b), (c) and (d) correspond to those highlighted in Fig.~\ref{fig:tree}.
    }\label{fig:mutations}
\end{figure}

The composability of motifs makes it intuitive to construct methods for generating new algorithms from a pair of parent algorithms. In particular, we used a `chop-and-join' mutation which separates each of a pair of algorithms into two pieces, where the division is made at a random point in the sequence of motifs. The first pieces of each algorithm are exchanged to produce two new algorithms. In addition, three types of mutations were used that operate on a single algorithm. The `property' mutation selects a motif at random from the composition of motifs and either modifies a randomly chosen property or replaces the selected motif with a randomly generated one. The `dropout' and `insert' mutations work by removing and inserting a randomly generated motif, respectively. The probability of using a mask, unmask, cycle, or pivot when randomly generating a motif is another hyper-parameter of the search. If the task uses an oracle then it is treated as a motif and is also assigned a probability of being used. Fig.~\ref{fig:mutations} shows an example of each of these four mutations. Each example corresponds to a mutation that occurred during the experiment discussed in Fig.~\ref{fig:tree}. If the chop-and-join mutation was omitted then such an evolutionary tree could only be a linear sequence of algorithms.

The process of selection and mutation is repeated according to the batch size hyper-parameter $n_{batch}$, as shown in Alg.~\ref{alg:genetic} lines~\ref{line:batch}-\ref{line:mutation}. The number of new algorithms created in each generation is $8n_{batch}$ since two algorithms are selected and we apply four mutations. The new candidate algorithms are evaluated in parallel and the batch size is chosen according to available hardware. Consequently, the rate of growth of the population is determined by the batch size and this should be taken into account when scheduling the probability of performing random or tournament selection. Once evaluated, the new algorithms are added to the population, see line~\ref{line:eval} in Alg.~\ref{alg:genetic}. 

Since the learning tasks that we tested were known algorithms, we implemented early stopping in  line~\ref{line:stop} Alg.~\ref{alg:genetic}. This is not a requirement of the method and line~\ref{line:stop} can be replaced with a tautology, creating an infinite loop. In our implementation, the primary stopping condition is met when the overlap between the state output by the algorithm and the target state is greater than a chosen threshold for all training examples. In addition, for problems that employ an oracle, we require that the composition of motifs contains at most one oracle call. Finally, we set a maximum runtime for an experiment. In general, it is useful to strike a balance between allowing an experiment to run for sufficient time and running multiple experiments. Performing multiple experiments is useful since the search strategy depends on several random processes which affect convergence.

\subsection{Performance estimation}\label{sec:fitness}

The fitness function associates a numerical value to the performance of a candidate algorithm. The fitness is then used to generate a relative ranking of the algorithms in the population. Designing a fitness function is important as it governs how the search space is navigated. We determine the fitness of a candidate algorithm by analysing its performance with respect to a set of training examples and by considering its complexity.  

The first contribution to our fitness function is constructed using state overlap. The training set, $T = \cup_{q\in {\mathcal{N}}} \{(x^{(q)}_i, y^{(q)}_i)\}_{i \in I}$, is a collection of input-target pairs, $(x^{(q)}_i, y^{(q)}_i)$, which are grouped by $q\in {\mathcal{N}}$, where ${\mathcal{N}}$ is a set of circuit sizes. We define the fitness function as follows
\begin{align}
f({\mathcal{A}}) &:= \min_{q\in {\mathcal{N}}} f_q({\mathcal{A}}) \\ &:= \min_{q\in {\mathcal{N}}} \left\{ \sum_{i\in I} |\braket{y^{(q)}_i|{U([1,\dots,q],\mathcal{A})}|x^{(q)}_i}|^2 \right\}\,,
\end{align}
where {$U([1,\dots,q],\mathcal{A})$ (Eq.~\eqref{eq:hypergraph_build_state})} is the quantum circuit (unitary operation) associated with the candidate algorithm ${\mathcal{A}}$ instantiated for the relevant circuit size implied by $q$. Here, we take the minimum fitness across the set of circuit sizes, but taking the mean is also valid. It turns out that it is sufficient to evaluate an algorithm on a small number of circuit sizes for the general structure to be learnt. In particular, circuits of size $2$, $3$ and $4$ qubits were sufficient for the Deutsch-Jozsa algorithm and QFT, and circuits of size $3$, $4$ and $5$ qubits were sufficient for Grover's search. 

The fitness can be improved by adding penalties to $f_q({\mathcal{A}})$ that depend on the complexity of ${U([1,\dots,q],\mathcal{A})}$. To favour smaller simpler structures, a penalty is added in terms of the number of unitary gates in the circuit ${U([1,\dots,q],\mathcal{A})}$. A penalty is also added for the number of oracle calls and the number of gate parameters, where applicable. Lastly, we add a penalty based on the Jensen–Shannon divergence~\cite{lin1991divergence} which penalises circuits that perform non-uniformly across the elements of the training set. The relative weights of these penalties are a hyper-parameter of the search.

When an algorithm contains parametrised gates, these are trained numerically using gradient descent
~\cite{paszke2017automatic, bergholm2018pennylane} to maximise the fitness of the candidate. In order to speed up evaluation, we allow parameter training to run for only a few seconds, which further penalises highly parameterised circuits. Parameters are later trained more carefully during post-processing.

During evaluation, the repetition parameter is increased incrementally from $1$ until the fitness no longer increases, or until the number of repetitions is equal to the circuit size. This upper limit on the number of repetitions is motivated by the fact that we would like the size of our learnt algorithm to scale at most linearly in the number of qubits. During post-processing, the relationship between the circuit size and the number of repetitions can be inferred by evaluating the algorithm on a larger number of circuit sizes.

\subsection{Tasks}\label{sec:tasks}

In our approach, setting up the problem is the primary task of the user, while the task of designing the algorithm is delegated to the computer. As we use a bottom-up approach, the user needs to carefully construct training examples that capture the desired behaviour of the algorithm. If required, the user should also define an appropriate oracle. Here, we simply used the known oracles. However, many formulations of the problem may need to be tested to arrive at the final algorithm, which is straightforward to do using our method. Below, we discuss the problem setup for each task. 

The quantum Fourier transform (QFT)~\cite{shor1999polynomial} maps an input state to an output state whose amplitudes are related by the discrete Fourier transform. The training examples are constructed using the computational basis states and their corresponding Fourier-transformed counterparts. An additional training pair is added where the input state is given by an equal superposition of all computational basis states. This penalises erroneous relative phase shifts between all other output states. 

The Deutsch-Jozsa algorithm~\cite{deutsch1992rapid} distinguishes between a constant and a balanced function, $f: \mathbb{R}\rightarrow \{0,1\}$, in a single function evaluation. An oracle is provided that performs the transformation $\ket{x}\ket{0} \rightarrow \ket{x}\ket{f(x)}$. Since we would like the learnt algorithm to use as few oracle calls as possible, i.e. only one, we assume at most a single repetition of the learnt algorithm.  The training set contains a training pair for each possible balanced function for the given circuit size. For each of these training pairs, we add another pair for the constant function, which makes the training set balanced. We distinguish between the training pairs associated with balanced and constant functions by specifying that the expected fidelity with respect to the zero state, $|\braket{y^{(q)}_i = 0 |{U([1,\dots,q,\mathcal{A}])}|x^{(q)}_i}|^2$, should be zero and one, respectively. 

Grover's search~\cite{grover1997quantum} provides an optimal method for searching an unstructured database. The oracle marks the state associated with the sought database entry with a phase, namely $\exp(i\pi)=-1$. A training pair is constructed for each of the possible sought entries (computational basis states) for the given problem size $2^q$, where $q$ is the number of qubits. A fixed input state of $(\frac{\ket{0}+\ket{1}}{\sqrt{2}})^{\otimes q}$ is used, which encodes the assumption that each database entry has an equal probability of being the sought entry. 

Oracles are implemented as Python lambda functions so that relevant parameters can be passed at the time of evaluation. Namely, the target state, in the case of Grover's search algorithm, and the function $f$ in the case of the Deutsch-Jozsa algorithm.

\section{Data Availability}

The dataset generated and analysed during the current study is not publicly available due to its size but is available from the corresponding author upon reasonable request.

\section{Code Availability}

The underlying code for this study can be accessed at \url{https://github.com/AmyRouillard/qalgosynth}.

\bibliographystyle{unsrtnat}
\bibliography{bib}

\section{Acknowledgements}

The computations were performed at University of Geneva using Baobab HPC service. AR acknowledges the financial assistance of the National Research Foundation (NRF) of South Africa. Opinions expressed and conclusions arrived at, are those of the author and are not necessarily to be attributed to the NRF. ML acknowledges the financial support of the Oppenheimer Memorial Trust of South Africa.  ML also acknowledges the financial support of the Unitary Foundation for the development of the Hierarqcal Python package. The funder played no role in study design, data collection, analysis and interpretation of data, or the writing of this manuscript. 

\section{Author Contributions}

AR and ML conceived and designed the study. AR implemented the search strategy, performed the numerical experiments and analysed the results. AR prepared the manuscript with input from ML. FP supervised the research. All authors read and approved the final manuscript.

\section{Competing interests}

All authors declare no financial or non-financial competing interests. 

\appendix
\section{{Introduction to the DSL}}\label{app:DSL}

A motif $M_m$ is a building block of the DSL and is used to generate a hypergraph $\mathcal{G}_m=(Q,E_m)$, associated with a unitary operation $U^{(m)}$. A circuit diagram can be interpreted as a visual representation of a hypergraph called a parallel aggregated ordered hypergraph (PAOH)~\cite{valdivia_analyzing_2021}. PAOH represents nodes as parallel horizontal lines, and edges as vertical lines, using dots to depict the connections to one or more nodes. Comparing this to a quantum circuit diagram, we can associate qubits with nodes and unitary gates with edges. Using the typical visual conventions for depicting quantum gates provides additional information but does not change the interpretation of the circuit as a PAOH. 

Each motif type corresponds to a different method of edge generation. Motifs can be composed to create a sequence of motifs, which can in turn be composed to form higher-level motifs, and so on. Composition is discussed in Sec.~\ref{app:composition} and forms the basis for building candidate algorithms $\mathcal{A}$, which are a sequence of $p$ motifs, $(M_1,M_2,\dots, M_p)$. For the node (qubit) indices $Q$, the quantum circuit is generated according to Eq.~\eqref{eq:hypergraph_build_state}. The order of gate execution matters for quantum circuits, and gates are added to the circuit from left to right as edges are generated. The importance of edge ordering differentiates the circuit representation from the PAOH. Below, we provide specific examples that illustrate the functionality of each motif in the DSL.

\subsection{Cycle}

\begin{figure}[ht]
    \begin{subfigure}{\columnwidth}
        \centering
        \begin{tcolorbox}[colback=white, colframe=black, boxrule=0.5mm, arc=0mm, left=2mm, right=2mm, top=2mm, bottom=2mm]
    \begin{minted}[fontsize=\footnotesize]{python}
Qcycle(
    mapping=Toffoli,
    stride=1,
    step=1,
    offset=0,
    boundary="periodic",
    edge_order=[1],
    share_weights = False,
)
    \end{minted}
    \end{tcolorbox}
    \end{subfigure}
    \\
    \begin{subfigure}{\columnwidth}
        \centering
        \includegraphics[height=0.7\textwidth]{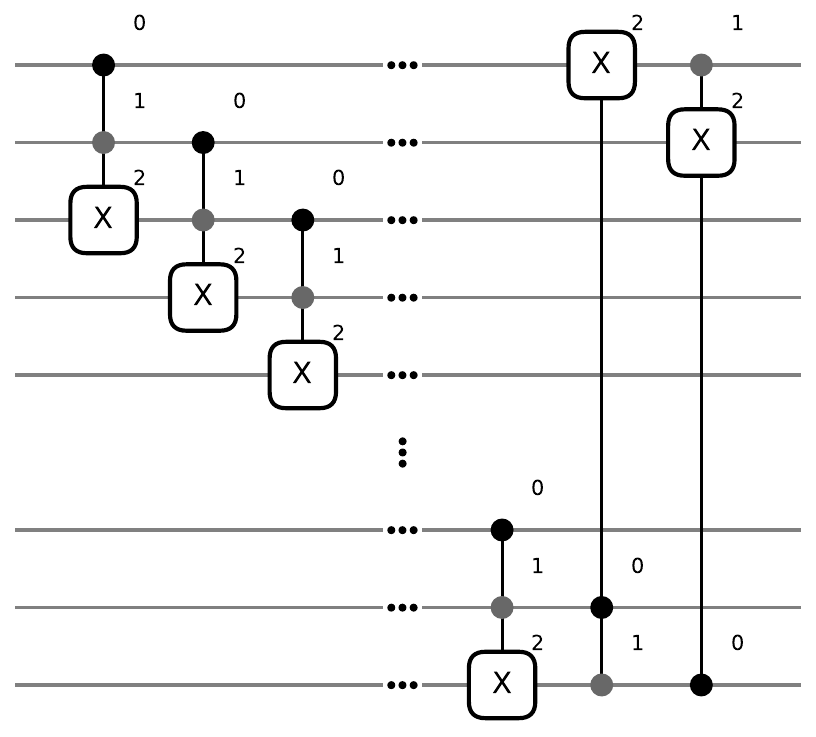}
    \end{subfigure}
    \caption[Default behaviour of the cycle]{Default behaviour of the cycle: The DSL motif \mintinline{python}|QCycle| (top) produces the quantum circuit (bottom). The mapping used is the Toffoli gate. The gate inputs and outputs are labelled $0$, $1$, $2$ to illustrate how the ordering is preserved when the cycle of gates wraps around from the bottom to the top of the circuit. }
    \label{fig:cycle_default}
\end{figure}

\begin{figure}[h]
    \begin{subfigure}{\columnwidth}
        \centering
        \begin{tcolorbox}[colback=white, colframe=black, boxrule=0.5mm, arc=0mm, left=2mm, right=2mm, top=2mm, bottom=2mm]
    \begin{minted}[fontsize=\footnotesize]{python}
Qcycle(
    mapping=Toffoli,
    stride=1,
    step=1,
    offset=0,
    boundary="open",
    edge_order=[1],
    share_weights = False,
)
    \end{minted}
    \end{tcolorbox}
    \end{subfigure}
    \\
    \begin{subfigure}{\columnwidth}
        \centering
        \includegraphics[height=0.7\textwidth]{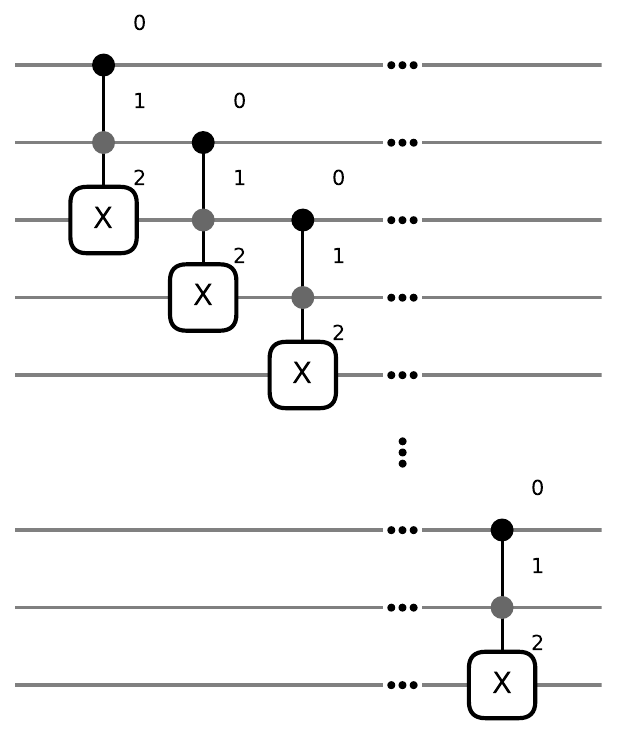}
    \end{subfigure}
    \caption[Cycle with open boundary condition]{Cycle with open boundary condition: The DSL motif \mintinline{python}|QCycle| (top) produces the quantum circuit (bottom). Compared to Fig.~\ref{fig:cycle_default}, the gates no longer wrap around from the bottom to the top of the circuit as a result of the open boundary condition.}
    \label{fig:cycle_open}
\end{figure}

As the name suggests, the cycle, represented as \mintinline{python}|QCycle| in the DSL, is a motif that generates a sequence of gates by cycling through the available qubits. Fig.~\ref{fig:cycle_default} shows the default behaviour of the cycle when used with a $3$-qubit gate. The cycle has a \textit{mapping} parameter that defines the operation associated with each edge. In this example, the mapping used is the Toffoli gate. A mapping can also be a parametrised gate, such as a parametrised rotation or phase gate. The list of edges generated by the cycle shown in Fig.~\ref{fig:cycle_default} is $E=[(1,2,3),(2,3,4),\dots,(q-2,q-1,q),(q-1,q,1),(q,1,2)]$, where $q$ is the number of qubits. By default, the cycle uses periodic boundary conditions, so that a unitary gate's inputs are assigned to qubits modulo $q$. Fig.~\ref{fig:cycle_open} shows the behaviour of the cycle when these boundary conditions are changed from \textit{periodic} to \textit{open}. For open boundary conditions, the list of edges is given by $E=[(1,2,3),(2,3,4),\dots,(q-2,q-1,q)]$.

\begin{figure}[h]
    \begin{subfigure}{\columnwidth}
        \centering
        \begin{tcolorbox}[colback=white, colframe=black, boxrule=0.5mm, arc=0mm, left=2mm, right=2mm, top=2mm, bottom=2mm]
    \begin{minted}[fontsize=\footnotesize]{python}
Qcycle(
    mapping=Toffoli,
    stride=1,
    step=1,
    offset=1,
    boundary="periodic",
    edge_order=[1],
    share_weights = False,
)
    \end{minted}
    \end{tcolorbox}
    \end{subfigure}
    \\
    \begin{subfigure}{\columnwidth}
        \centering
        \includegraphics[height=0.7\textwidth]{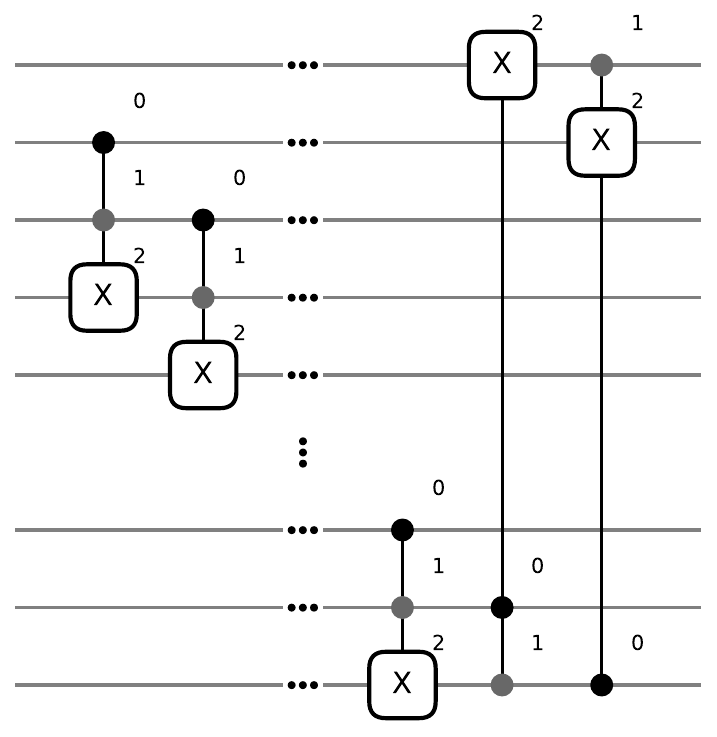}
    \end{subfigure}
    \caption[Cycle with offset]{Cycle with offset: The DSL motif \mintinline{python}|QCycle| (top) produces the quantum circuit (bottom). The gates are placed with an offset of $1$ qubit with respect to their default position, shown in Fig.~\ref{fig:cycle_default}.}
    \label{fig:cycle_offset}
\end{figure}

The \textit{stride}, \textit{step}, and \textit{offset} parameters take on integer values and control how the unitary gate inputs are assigned to qubits, i.e., how edges are generated. The offset controls where the assignment begins within the available qubits. An offset of $1$ effectively shifts all gates down by one qubit, as demonstrated in Fig.~\ref{fig:cycle_offset}. Here, the list of edge is given by $E=[(2,3,4),(3,4,5),\dots,(q-2,q-1,q),(q-1,q,1),(q,1,2)]$. Adding an offset does not necessarily preserve the number of gates (edges) generated for two circuits of the same size. For example, Fig.~\ref{fig:cycle_default} has one more gate than Fig.~\ref{fig:cycle_offset}.

\begin{figure}[h]
    \begin{subfigure}{\columnwidth}
        \centering
        \begin{tcolorbox}[colback=white, colframe=black, boxrule=0.5mm, arc=0mm, left=2mm, right=2mm, top=2mm, bottom=2mm]
    \begin{minted}[fontsize=\footnotesize]{python}
Qcycle(
    mapping=Toffoli,
    stride=1,
    step=2,
    offset=0,
    boundary="periodic",
    edge_order=[1],
    share_weights = False,
)
    \end{minted}
    \end{tcolorbox}
    \end{subfigure}
    \\
    \begin{subfigure}{\columnwidth}
        \centering
        \includegraphics[height=0.7\textwidth]{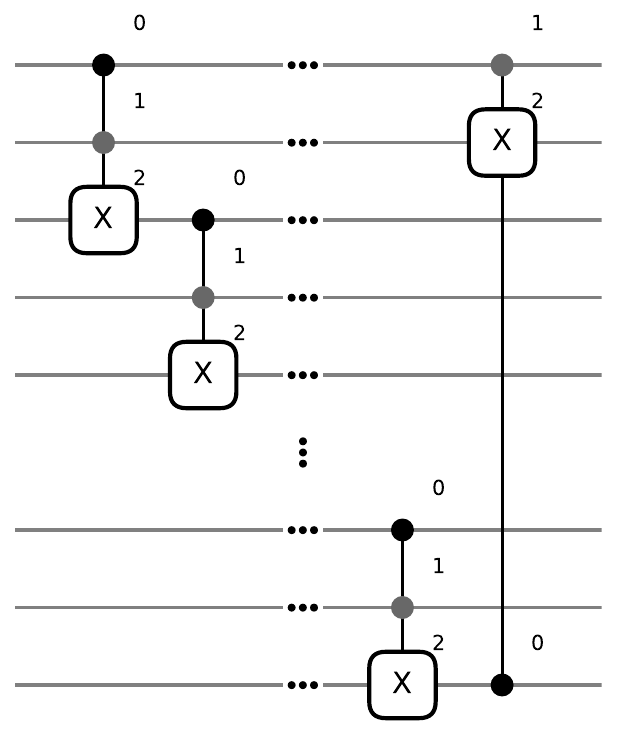}
    \end{subfigure}
    \caption[Changing the step of the cycle]{Changing the step of the cycle: The DSL motif \mintinline{python}|QCycle| (top) produces the quantum circuit (bottom). The step parameter controls which qubits become the first node in each edge. Changing the step from $1$ to $2$ results in a gate being generated at every second qubit. Increasing the step to $3$ would result in a gate at every third qubit, and so on.}
    \label{fig:cycle_step}
\end{figure}

The \textit{step} parameter, like the offset, controls the traversal of the available qubits. A step of $1$ results in every qubit being visited, if we ignore the effect of the offset. While a step of $n$ results in every $n$th qubit being visited, starting from the first (plus the offset) and proceeding to the last. Fig.~\ref{fig:cycle_step} illustrates the effect of changing the default step to a value of $2$. In this example, the list of edges is given by $E =[(1,2,3),(3,4,5),\dots,(q-2,q-1,q),(q,1,2)]$.

\begin{figure}[h]
    \begin{subfigure}{\columnwidth}
        \centering
        \begin{tcolorbox}[colback=white, colframe=black, boxrule=0.5mm, arc=0mm, left=2mm, right=2mm, top=2mm, bottom=2mm]
    \begin{minted}[fontsize=\footnotesize]{python}
Qcycle(
    mapping=Toffoli,
    stride=2,
    step=1,
    offset=0,
    boundary="periodic",
    edge_order=[1],
    share_weights = False,
)
    \end{minted}
    \end{tcolorbox}
    \end{subfigure}
    \\
    \begin{subfigure}{\columnwidth}
        \centering
        \includegraphics[height=0.7\textwidth]{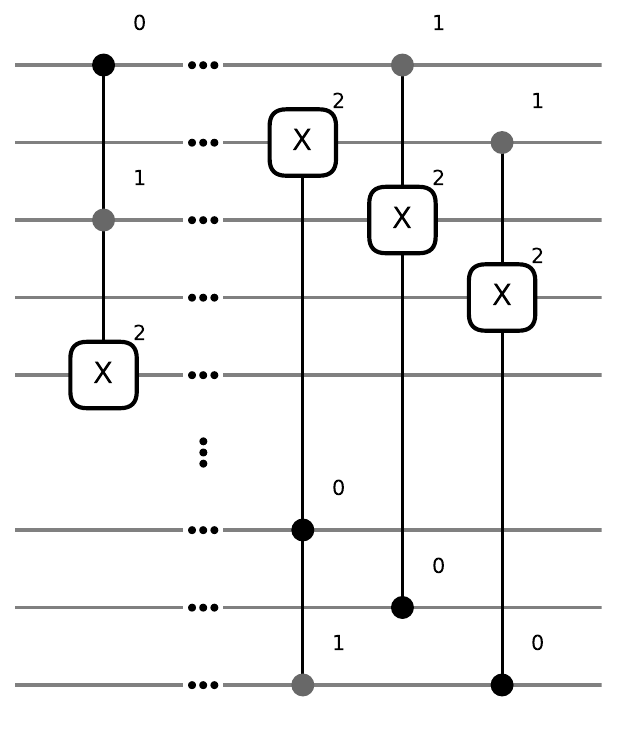}
    \end{subfigure}
    \caption[Changing the stride of the cycle]{Changing the stride of the cycle: The DSL motif \mintinline{python}|QCycle| (top) produces the quantum circuit (bottom). Increasing the stride increases the distance between nodes connected by the same edge. 
    }
    \label{fig:cycle_stride}
\end{figure}

The \textit{stride} controls the interval between node indices connected by the same edge, that is, the intervals between qubit indices assigned to the same unitary gate. For example, a stride of $1$, as in the default case, results in edges being generated for qubit indices that are a distance of $1$ apart. Increasing the stride results in an increased uniform distance between qubit index assignments. Fig.~\ref{fig:cycle_stride} illustrates the gate placement for a stride of $2$ and a step of $1$. In this example, the list of edges is given by $E=[(1,3,5),(2,4,6),\dots,(q-2,q,2),(q-1,1,3),(q,2,4)]$.

\begin{figure}[h]
    \begin{subfigure}{\columnwidth}
        \centering
        \begin{tcolorbox}[colback=white, colframe=black, boxrule=0.5mm, arc=0mm, left=2mm, right=2mm, top=2mm, bottom=2mm]
    \begin{minted}[fontsize=\footnotesize]{python}
Qcycle(
    mapping=Toffoli,
    stride=1,
    step=1,
    offset=0,
    boundary="periodic",
    edge_order=[-1],
    share_weights = False,
)
    \end{minted}
    \end{tcolorbox}
    \end{subfigure}
    \\
    \begin{subfigure}{\columnwidth}
        \centering
        \includegraphics[height=0.7\textwidth]{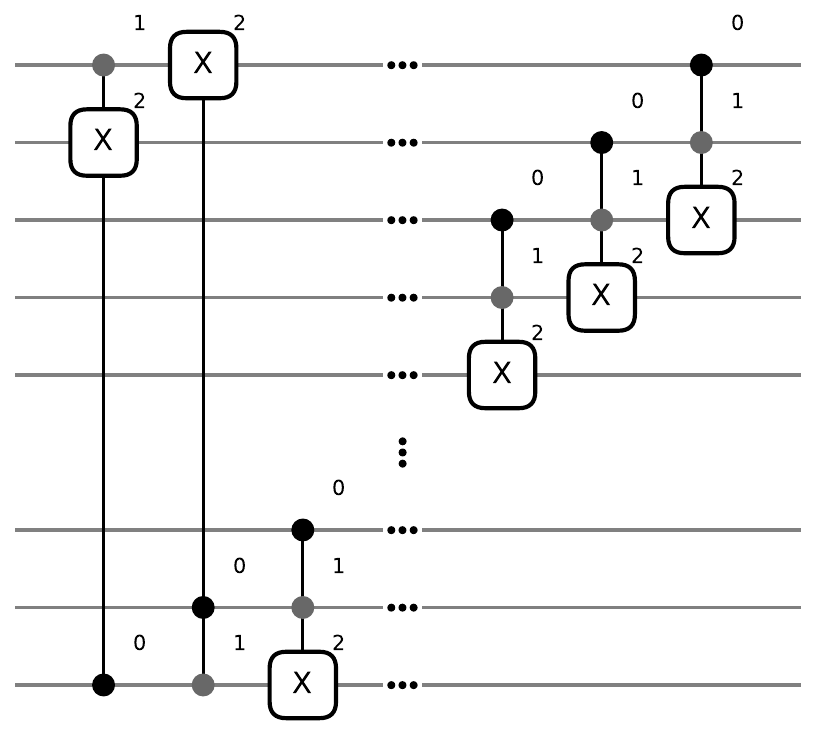}
    \end{subfigure}
    \caption[Cycle with reversed edge order]{Cycle with reversed edge order: The DSL motif \mintinline{python}|QCycle| (top) produces the quantum circuit (bottom). Compared to Fig.~\ref{fig:cycle_default}, the gates appear in reverse order.}
    \label{fig:cycle_edge}
\end{figure}

The \textit{edge order} parameter controls the order of the gates (edges) in the sequence. The order can be specified using a list of integers or by using one of the two special inputs: \mintinline{python}|[1]| indicating default ordering, and \mintinline{python}|[-1]| indicating reverse ordering. Fig.~\ref{fig:cycle_edge} demonstrates the reverse ordering functionality, where the list of edges is given by $E=[(q,1,2),(q-1,q,1),(q-2,q-1,q),\dots,(2,3,4),(1,2,3)]$.

Based on these examples, it is intuitive to understand how gate (edge) generation will change when the unitary mapping used is a one-, two-, or $n$-qubit gate. Notably, changing the stride when a one-qubit gate is used does not affect the gate placement. Each edge connects to a single qubit (node) since a one-qubit gate has a single input. In this case, edge generation is only affected by the offset and step parameters.

Finally, the \textit{share weights} Boolean parameter sets all gate parameters identically if set to true. This can significantly reduce the number of trainable parameters in the circuit. For non-parametric gates such as the Toffoli gate, changing \textit{share weights} does not have an effect.

\subsection{Pivot}

The pivot motif, implemented as \mintinline{python}|Qpivot| in the DSL, is so named because it connects a set of \textit{pivot} qubits to the set of remaining available qubits. Like the cycle, the pivot has a \textit{mapping} parameter that defines the operation associated with each edge. Pivot qubits are selected using the \textit{global pattern} parameter. The \textit{merge within} parameter controls which gate inputs are assigned to qubits in the pivot set, with the remainder being assigned to non-pivot qubits. Both of these parameters make use of bit strings, where a $1$ indicates a pivot qubit. 

\begin{figure}[h]
    \begin{subfigure}{\columnwidth}
        \centering
        \begin{tcolorbox}[colback=white, colframe=black, boxrule=0.5mm, arc=0mm, left=2mm, right=2mm, top=2mm, bottom=2mm]
    \begin{minted}[fontsize=\footnotesize]{python}
Qpivot(
    mapping=Toffoli,
    global_pattern="1*",
    merge_within="1*",
    strides=[1, 1, 0],
    steps=[1, 1, 1],
    offsets=[0, 0, 0],
    boundaries=["open", "open", "periodic"],
    edge_order=[1],
)
    \end{minted}
    \end{tcolorbox}
    \end{subfigure}
    \\
    \begin{subfigure}{\columnwidth}
        \centering
        \includegraphics[height=0.7\textwidth]{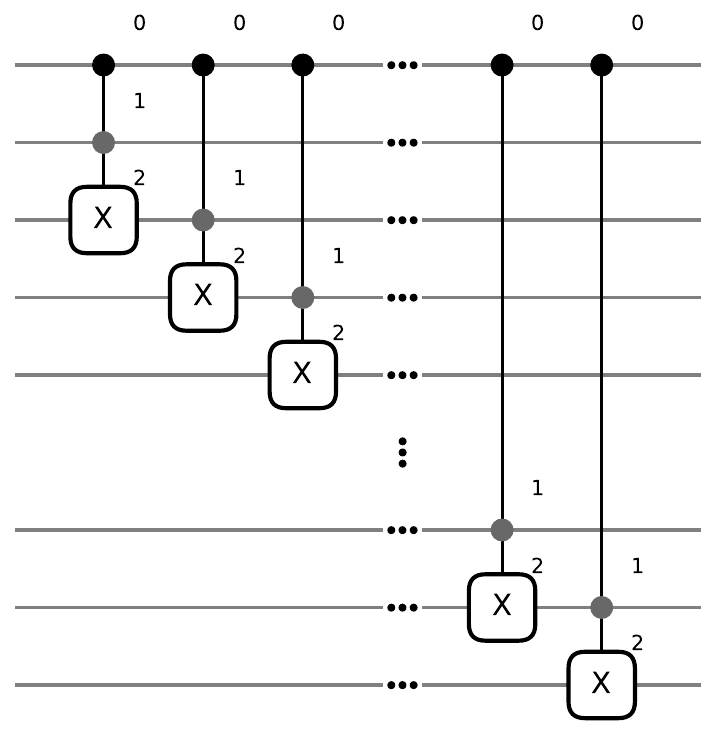}
    \end{subfigure}
    \caption[Default behaviour of the pivot]{Default behaviour of the pivot: The DSL motif \mintinline{python}|Qpivot| (top) produces the quantum circuit (bottom). The \textit{global pattern} sets the first qubit in the circuit as the pivot qubit. Each gate is placed by cycling through the remaining qubits while making a fixed assignment of the first input of the gate to the top qubit, as specified by the \textit{merge within} pattern. }
    \label{fig:pivot_default}
\end{figure}

\begin{figure}[H]
    \begin{subfigure}{\columnwidth}
        \centering
        \begin{tcolorbox}[colback=white, colframe=black, boxrule=0.5mm, arc=0mm, left=2mm, right=2mm, top=2mm, bottom=2mm]
    \begin{minted}[fontsize=\footnotesize]{python}
Qpivot(
    mapping=Toffoli,
    global_pattern="*1",
    merge_within="1*",
    strides=[1, 1, 0],
    steps=[1, 1, 1],
    offsets=[0, 0, 0],
    boundaries=["open",  "open", "periodic"],
    edge_order=[1],
)
    \end{minted}
    \end{tcolorbox}
    \end{subfigure}
    \\
    \begin{subfigure}{\columnwidth}
        \centering
        \includegraphics[height=0.7\textwidth]{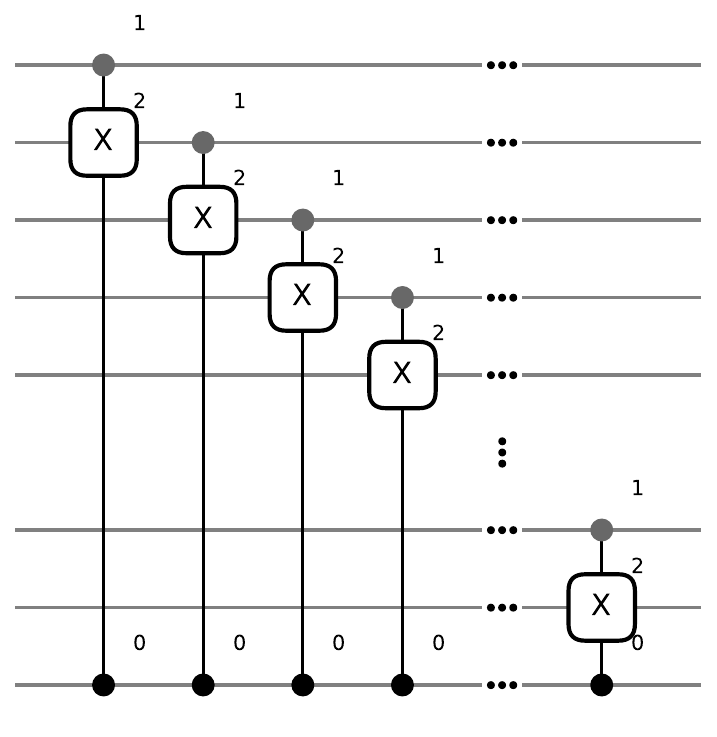}
    \end{subfigure}
    \caption[Controlling the pivot qubits using the global pattern]{Controlling the pivot qubits using the \textit{global pattern}: The DSL motif \mintinline{python}|Qpivot| (top) produces the quantum circuit (bottom). Instead of pivoting on the first qubit, as seen in Fig.~\ref{fig:pivot_global_pattern}, we now pivot on the last qubit in the circuit. This results in the last qubit being assigned to the first input of each generated gate.}
    \label{fig:pivot_global_pattern}
\end{figure}

The DSL is also equipped with the following useful \textit{wildcard} conventions: a star, \mintinline{python}|*|, is filled with zeros, while an exclamation mark, \mintinline{python}|!|, is filled with ones. For example, as in Fig.~\ref{fig:pivot_default}, a \textit{global pattern} of \mintinline{python}|1*| is interpreted as \mintinline{python}|100...0|, so that the motif \textit{pivots} on the first qubit in the circuit. The remaining gate inputs are assigned to available qubits not in the pivot set, following the same edge generation rules as the cycle. By setting the \textit{merge within} parameter to \mintinline{python}|1*|, the first gate input is assigned to the pivot qubit(s). Changing the \textit{global pattern} from \mintinline{python}|1*| to \mintinline{python}|*1| results in a pivot on the last qubit in the circuit, as shown in Fig.~\ref{fig:pivot_global_pattern}. In addition, a string not containing \mintinline{python}|*| or \mintinline{python}|!| will be assumed to be repeating. For example, \mintinline{python}|01| is expanded to the required length as \mintinline{python}|010101...|, and truncated from the right if necessary. 

The behaviour of the pivot can be specified more precisely as follows. The \textit{merge within} parameter subdivides a unitary gate's inputs into two sets: those that will be assigned to pivot qubits and those that will be assigned to the rest. The \textit{global pattern} similarly subdivides the available qubits (nodes) into pivot and non-pivot qubits. Edges are generated for each of these subsets using the cycle generation pattern. Let us call these the pivot cycle and rest cycle, respectively. Edge generation is completed by merging the two cycles, where this merger occurs using the same logic as used to generate a cycle. For this reason, the pivot requires three values for the \textit{strides}, \textit{steps}, \textit{offsets}, and \textit{boundaries} parameters. These three values control the behaviour of the pivot cycle, rest cycle, and merger, respectively. During the merger, we step through the edges associated with the rest cycle and join them to edges from the pivot cycle, creating edges that connect the appropriate number of nodes. For example, in Fig.~\ref{fig:pivot_default}, where we pivot on the first qubit, the pivot cycle has edge list $E_p=[(1)]$ and the rest cycle an edge list $E_r=[(2,3),(3,4),\dots,(q-2,q-1),(q-1,q)]$. These are merged into the final edge list $E=[(1,2,3),(1,3,4),\dots,(1,q-2,q-1),(1,q-1,q)]$. While in Fig.~\ref{fig:pivot_global_pattern}, where we pivot on the last qubit, the pivot cycle has edge list $E_p=[(q)]$ and the rest cycle an edge list $E_r=[(1,2),(2,3),\dots,(q-2,q-1)]$. These are combined into the final edge list $E=[(q,1,2),(q,2,3),\dots,(q,q-2,q-1)]$.

\begin{figure}[H]
    \begin{subfigure}{\columnwidth}
        \centering
        \begin{tcolorbox}[colback=white, colframe=black, boxrule=0.5mm, arc=0mm, left=2mm, right=2mm, top=2mm, bottom=2mm]
    \begin{minted}[fontsize=\footnotesize]{python}
Qpivot(
    mapping=Toffoli,
    global_pattern="11*",
    merge_within="1*",
    strides=[1, 1, 0],
    steps=[1, 1, 1],
    offsets=[0, 0, 0],
    boundaries=["open", "open", "periodic"],
    edge_order=[1],
)
    \end{minted}
    \end{tcolorbox}
    \end{subfigure}
    \\
    \begin{subfigure}{\columnwidth}
        \centering
        \includegraphics[height=0.7\textwidth]{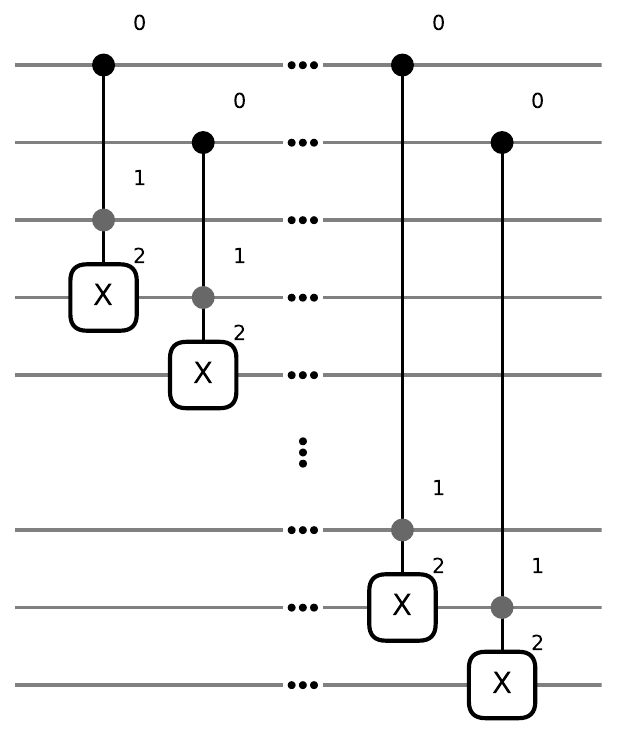}
    \end{subfigure}
    \caption[Pivoting on more than one qubit]{Pivoting on more than one qubit: The DSL motif \mintinline{python}|Qpivot| (top) produces the quantum circuit (bottom). In this example, there are two pivot qubits specified by the \textit{global pattern}. A first set of edges is generated by a cycle for a one-qubit gate on the two pivot qubits. A second set of edges is generated by a cycle for a two-qubit gate on the remaining available qubits. The final edges are generated by merging edges from these sets, where elements of the second set are merged with elements of the first set cyclically until all elements of the second set are exhausted.}
    \label{fig:pivot_global_pattern2}
\end{figure}

In Fig.~\ref{fig:pivot_global_pattern2}, two pivot qubits have been specified and the merge within pattern is \mintinline{python}|1*|. Therefore, a rest cycle is generated for an effective two-qubit gate (second and third input of the Toffoli gate) on the subset of non-pivot qubits, with edge list $E_r=[(3,4),(4,5),\dots,(q-2,q-1),(q-1,q)]$. The pivot cycle is generated for an effective one-qubit gate (first input of the Toffoli gate) on the two selected pivot qubits, with edge list $E_p=[(1),(2)]$. These are merged using periodic boundary conditions to produce the final edge list $E=[(1,3,4),(2,4,5),\dots,(1,q-2,q-1),(2,q-1,q)]$.

\begin{figure}[H]
    \begin{subfigure}{\columnwidth}
        \centering
        \begin{tcolorbox}[colback=white, colframe=black, boxrule=0.5mm, arc=0mm, left=2mm, right=2mm, top=2mm, bottom=2mm]
    \begin{minted}[fontsize=\footnotesize]{python}
Qpivot(
    mapping=Toffoli,
    global_pattern="1*",
    merge_within="1*",
    strides=[1, 1, 0],
    steps=[1, 2, 1],
    offsets=[0, 0, 0],
    boundaries=["open", "open", "periodic"],
    edge_order=[1],
)
    \end{minted}
    \end{tcolorbox}
    \end{subfigure}
    \\
    \begin{subfigure}{\columnwidth}
        \centering
        \includegraphics[height=0.7\textwidth]{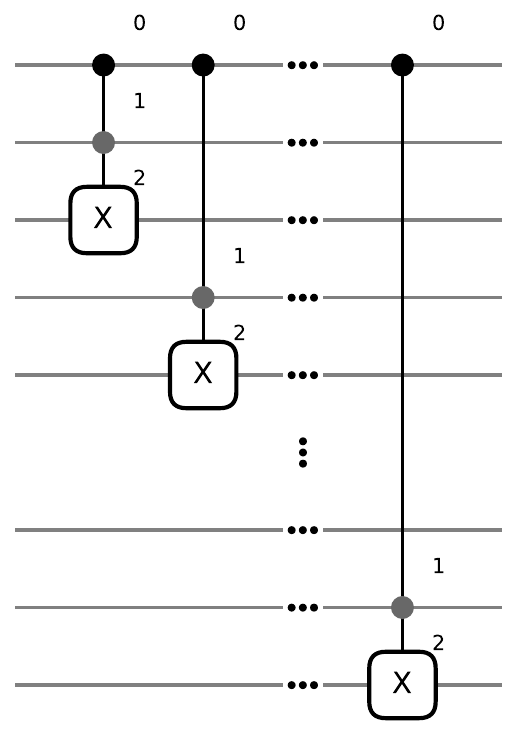}
    \end{subfigure}
    \caption[Additional pivot controls]{Additional pivot controls: The DSL motif \mintinline{python}|Qpivot| (top) produces the quantum circuit (bottom). The behaviour of the pivot cycle, rest cycle, and merger is controlled by the elements of the \textit{steps}, \textit{strides}, \textit{offsets}, and \textit{boundaries} parameters. By setting the second element of \textit{steps} to $2$, the rest cycle now uses a step of $2$, compared to a step of $1$ in Fig.~\ref{fig:pivot_default}.}
    \label{fig:pivot_steps}
\end{figure}

Similarly, for a single pivot qubit, we can introduce a step of $2$ in the rest cycle, as shown in Fig.~\ref{fig:pivot_steps} by setting the second element of the \textit{steps} parameter to $2$. In this example, the pivot cycle has edge list $E_p=[(1)]$ and the rest cycle edge list $E_r=[(2,3),(4,5),\dots,(q-3,q-2),(q-1,q)]$, with final edge list $E=[(1,2,3),(1,4,5),\dots,(1,q-3,q-2),(1,q-1,q)]$.

To avoid unexpected behaviour, the number of $1$s in the \textit{global pattern} is systematically increased to match the number of $1$s in \textit{merge within}, if this is not already the case. For example, a \textit{merge within} pattern of \mintinline{python}|111| and a global pattern of \mintinline{python}|1*| are interpreted as \mintinline{python}|111| and \mintinline{python}|111*|, respectively. 

\begin{figure}[H]
    \begin{subfigure}{\columnwidth}
        \centering
        \begin{tcolorbox}[colback=white, colframe=black, boxrule=0.5mm, arc=0mm, left=2mm, right=2mm, top=2mm, bottom=2mm]
    \begin{minted}[fontsize=\footnotesize]{python}
Qpivot(
    mapping=Toffoli,
    global_pattern="0101*1",
    merge_within="!",
    strides=[1, 1, 0],
    steps=[1, 1, 1],
    offsets=[0, 0, 0],
    boundaries=["open", "open", "periodic"],
    edge_order=[1],
)
    \end{minted}
    \end{tcolorbox}
    \end{subfigure}
    \hfill
    \begin{subfigure}{\columnwidth}
        \centering
        \includegraphics[height=0.7\textwidth]{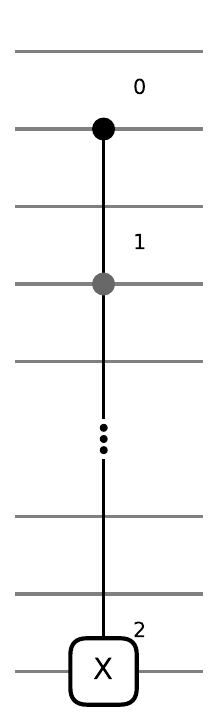}
    \end{subfigure}
    \caption[Using the pivot to place a single gate]{Using the pivot to place a single gate: The DSL motif \mintinline{python}|Qpivot| (top) produces the quantum circuit (bottom). By specifying a global pattern containing as many 1's as the size of the gate, a single gate can be placed on specific qubits. In this case, the three-qubit gate is connected to the second, fourth and last qubit. }
    \label{fig:pivot_place}
\end{figure}

Any single gate placement can be achieved using the pivot motif with an appropriate global pattern. This ensures that it is always possible to map an arbitrary gate-based circuit representation into the DSL. For example, as shown in Fig.~\ref{fig:pivot_place}, the pivot is used to place a single $3$-qubit gate at the second, fourth, and last qubit. The \textit{merge within} pattern is set to \mintinline{python}|!|, which encodes all $1$s. Therefore, all gate inputs are assigned to qubits from the pivot set and none elsewhere. Finally, the \textit{edge order} parameter changes the order of the gates, identically to the \textit{edge order} for the cycle. For this reason, an example is not shown.

\subsection{Motif composition}\label{app:composition}

The DSL allows motifs to be composed using both the plus and multiplication operations, as shown in Fig.~\ref{fig:composition}. The cycle of Hadamard gates is placed to the left of the cycle of controlled-not gates according to the order of operations in the composition of motifs. This entire structure is repeated three times by multiplying the composition by $3$. We can consider this a first example of a candidate algorithm, and it represents the typical way in which algorithms are constructed in the DSL.

\subsection{Mask and Unmask}

\begin{figure*}
    \begin{subfigure}{0.6\columnwidth}
        \centering
        \begin{tcolorbox}[colback=white, colframe=black, boxrule=0.5mm, arc=0mm, left=2mm, right=2mm, top=2mm, bottom=2mm]
    \begin{minted}[fontsize=\footnotesize]{python}
(
 Qcycle(mapping=H) 
 + Qcycle(mapping=CNOT)
) * 3


    \end{minted}
    \end{tcolorbox}
    \end{subfigure}
    \hspace{6pt}
    \begin{subfigure}{1.3\columnwidth}
        \centering
        \includegraphics[height=0.3\textwidth]{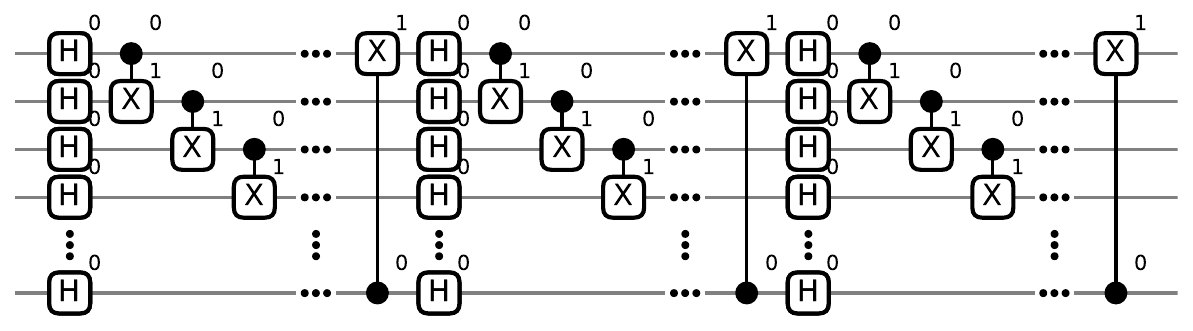}
    \end{subfigure}
    \caption[Composing motifs]{Composing motifs: On the left, two cycles are added together and the resulting motif multiplied by three. On the right is the equivalent circuit representation, in which a cycle of Hadamard gates is followed by a cycle of controlled-not operations, repeated three times. Where motif parameters are unspecified, they are assumed to use default values, as specified in Fig.~\ref{fig:cycle_default} and Fig.~\ref{fig:pivot_default}.}
    \label{fig:composition}
\end{figure*}

\begin{figure*}
    \begin{subfigure}{0.6\columnwidth}
        \centering
        \begin{tcolorbox}[colback=white, colframe=black, boxrule=0.5mm, arc=0mm, left=2mm, right=2mm, top=2mm, bottom=2mm]
    \begin{minted}[fontsize=\footnotesize]{python}
(
 Qcycle(mapping=H) 
 + Qcycle(mapping=CNOT)
 + Qmask(
    global_pattern="1*"
 )
) * 3
    \end{minted}
    \end{tcolorbox}
    \end{subfigure}%
    \hspace{6pt}
    \begin{subfigure}{1.3\columnwidth}
        \centering
        \includegraphics[height=0.3\textwidth]{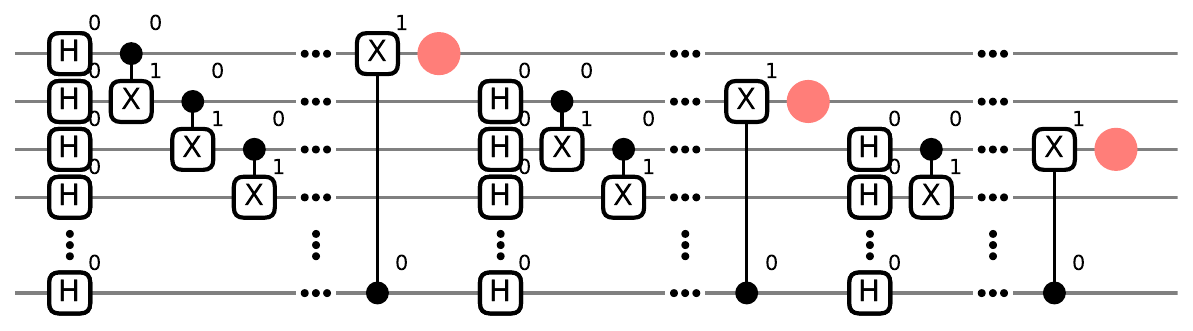}
    \end{subfigure}
    \caption[Composing motifs with mask operations]{Composing motifs with mask operations: On the left, two cycles and a mask are added together and repeated three times. This results in the circuit on the right, where, with each repetition, the first qubit in the circuit is made unavailable. Where motif parameters are unspecified, they are assumed to use default values, as specified in Fig.~\ref{fig:cycle_default} and Fig.~\ref{fig:pivot_default}.}
    \label{fig:composition_mask}
\end{figure*}

\begin{figure*}
    \begin{subfigure}{0.6\columnwidth}
        \centering
        \begin{tcolorbox}[colback=white, colframe=black, boxrule=0.5mm, arc=0mm, left=2mm, right=2mm, top=2mm, bottom=2mm]
    \begin{minted}[fontsize=\footnotesize]{python}
(
  Qcycle(mapping=H) 
  + Qcycle(mapping=CNOT)
  + Qmask(global_pattern="1*")
) * 3
+ (
  Qunmask(
    global_pattern="previous"
  )
) * 2
+ Qcycle(mapping=H)
    \end{minted}
    \end{tcolorbox}
    \end{subfigure}%
    \hspace{6pt}
    \begin{subfigure}{1.3\columnwidth}
        \centering
        \includegraphics[height=0.25\textwidth]{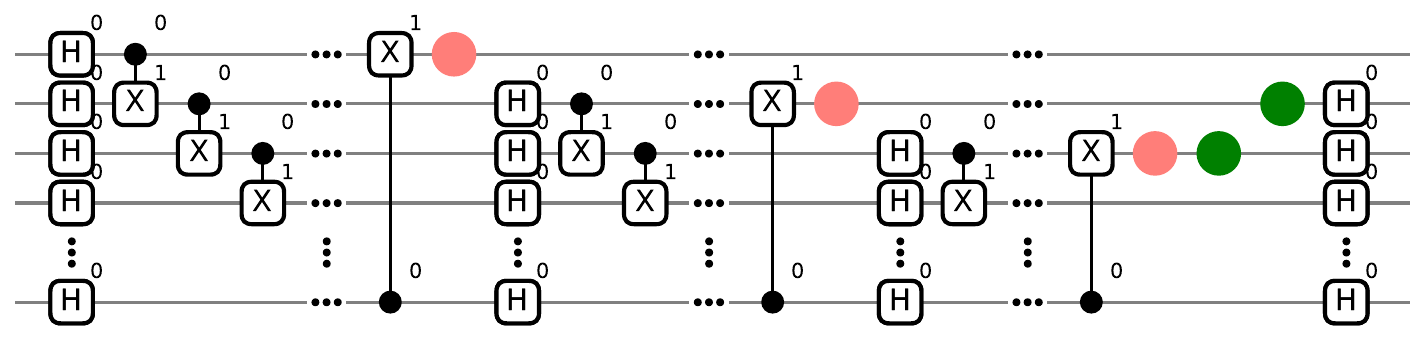}
    \end{subfigure}
    \caption[Unmasking previously masked qubits]{Unmasking previously masked qubits: Using the global pattern \mintinline{python}|"previous"|, we can undo the most recent mask operation to make qubits available again. Where motif parameters are unspecified, they are assumed to use default values, as specified in Figures~\ref{fig:cycle_default} and \ref{fig:pivot_default}.}
    \label{fig:composition_unmask}
\end{figure*}

The \mintinline{python}|Qmask| and \mintinline{python}|Qunmask| motifs serve to make qubits available and unavailable. It can be useful to make a subset of qubits temporarily unavailable, such as ancillary qubits. The mask and unmask also use a global pattern, which specifies which qubits are to be masked or unmasked, respectively. Fig.~\ref{fig:composition_mask} shows the behaviour of the mask and can be directly compared to Fig.~\ref{fig:composition}. The unmask has a special global pattern \mintinline{python}|"previous"| which undoes the operation of the previous mask, as shown in Fig.~\ref{fig:composition_unmask}.

\clearpage


\end{document}